\documentclass[pdftex,twocolumn,epjc3,amsmath,amssymb]{svjour3}          

\RequirePackage[T1]{fontenc}

\smartqed  

\RequirePackage{graphicx}
\RequirePackage{mathptmx}      
\RequirePackage{flushend}
\RequirePackage[numbers,sort&compress]{natbib}
\RequirePackage[colorlinks,citecolor=blue,urlcolor=blue,linkcolor=blue]{hyperref}
\RequirePackage{subfigure}
\RequirePackage{amssymb}
\hypersetup{pdfstartview=}

\journalname{Eur. Phys. J. C}

\begin{document}

\title{Anomalous magnetic and weak  magnetic dipole moments of the $\tau$ lepton in the simplest little Higgs model}



\author{M. A. Arroyo-Ure\~na    \thanksref{e1,addr1}     \and
        G. Hern\'andez-Tom\'e  \thanksref{e2,addr1}\and
        G. Tavares-Velasco  \thanksref{e3,addr1}
}

\thankstext{e1}{marcofis@yahoo.com.mx}
\thankstext{e2}{tome\_gerar@hotmail.com}
\thankstext{e3}{gtv@fcfm.buap.mx}

\institute{
              Facultad de Ciencias F\'isico-Matem\'aticas,\\
  Benem\'erita Universidad Aut\'onoma de Puebla,\\
 C.P. 72570, Puebla, Pue., Mexico \label{addr1}
}

\date{Received: date / Accepted: date}

\maketitle
\begin{abstract}
We obtain analytical expressions, both in terms of parametric integrals and   Passarino-Veltman scalar functions,  for the one-loop contributions to  the anomalous weak magnetic dipole moment (AWMDM) of a charged lepton in the framework of the simplest little Higgs model (SLHM). Our results are general and can be useful to compute the weak properties of a charged lepton in other extensions of the standard model (SM).  As a by-product we obtain generic  contributions to the anomalous magnetic dipole  moment (AMDM), which agree with previous results. We then study numerically the potential contributions from this model to the $\tau$ lepton  AMDM and AWMDM   for values of the parameter space   consistent with current experimental data. It is found that they  depend mainly  on the energy scale  $f$ at which the global symmetry is broken and the $t_\beta$ parameter, whereas there is little sensitivity to a mild change in the values of other  parameters of the model. While the $\tau$ AMDM is of the order of $10^{-9}$, the real (imaginary) part of its AWMDM is of the order of $10^{-9}$ ($10^{-10}$). These values seem to be out of the reach of the expected experimental sensitivity of future experiments.

\end{abstract}

\section{Introduction}
\label{intro}

 The most general dimension-five effective interaction of  a neutral $V$ gauge boson ($V=\gamma, Z$) to a charged lepton that respects Lorentz invariance   can be written in terms of six independent form factors:

\begin{eqnarray}
\Gamma_{\mu}^{\ell\ell V}(q^{2})&=&ie\Bigg[\gamma_{\mu}\left(F_{V}^{V}-F_{A}^{V}\gamma_{5}\right)-q_{\mu}\left(iF_{S}^{V}+F_{P}^{V}\gamma_{5}\right)
\nonumber\\&+&
\sigma_{\mu\nu}q^{\nu}\left(iF_{M}^{V}+F_{E}^{V}\gamma_{5}\right)\Bigg],
\end{eqnarray}
where $q$ is the incoming transfer four-momentum of the   gauge boson.  The electromagnetic (weak) properties of the lepton are determined by the photon ($Z$ gauge boson) vertex function. The $CP$-violating terms  define the static electric dipole moment (EDM) and the static weak electric dipole moment (WEDM):
\begin{eqnarray}
d_{\ell}     & = & -eF_{E}^{\gamma}(0),\\
d_{\ell}^{W} & = & -eF_{E}^{Z}(m_{Z}^{2}).
\end{eqnarray}
Since the standard model (SM) predictions for these $CP$-violating dipole moments are highly suppressed, they may serve to search  for new sources of $CP$ violation. In this work we are interested instead in the  static anomalous magnetic dipole moment (AMDM) and the static anomalous weak magnetic dipole moment (AWMDM), which are defined in terms of the $CP$-even form factor  as follows:
\begin{eqnarray}
a_{\ell}     & = & -2m_{\ell}F_{M}^{\gamma}(0),\\
a_{\ell}^{W} & = & -2m_{\ell}F_{M}^{Z}(m_{Z}^{2}).
\end{eqnarray}
The measurement of the AMDM of a lepton has long been considered  a probe for the SM, which considers leptons as point-like objects. The  AMDM of the electron $a_e$, which receives its main contributions  from quantum electrodynamics (QED),   has been calculated up to order of $\alpha^5$ \cite{Aoyama:2012wj} and the agreement between the theoretical  and the experimental values has reached the level of  ten significant digits \cite{Hanneke:2008tm}, which  represents one of the greatest milestones of QED.

 As far as the muon is concerned, the E821 experiment at Brookhaven National Lab (BNL) measured its AMDM $a_{\mu}$  with an  unprecedent precision of 0.54 ppm.  The current average experimental measurement is \cite{Bennett:2006fi}
 \begin{equation}
 a^{Exp.}_\mu = 11 659 209.1(5.4)(3.3)\times 10^{-10},
\end{equation}
and further  improvement is expected ats future experiments  by the $ (g-2)_\mu$ \cite{Kaspar:2015jwa} and    J-PARC $(g-2)$/EDM \cite{Saito:2012zz} Collaborations, which  aim to reach a precision of $\pm  0.2$ ppm. As for the theoretical prediction of the SM \cite{Olive:2016xmw}
\begin{equation}
a_{\mu}^{SM}=116591803(1)(42)(26)\times 10^{-11},
\end{equation}
there is still a large uncertainty in the estimate of the hadronic contribution, whereas the QED and electroweak contributions have been determined with a great precision \cite{Jegerlehner:2009ry}. Thus a more accurate evaluation of the leading order hadronic contribution  together with the future experimental measurement are needed to settle down the discrepancy between the SM prediction and the experimental value of $a_\mu$, which currently stands  at the level of 3.6 standard deviations \cite{Olive:2016xmw}:
\begin{equation}
\label{discrepancy}
\Delta a_{\mu}=a_{\mu}^{Exp.}-a_{\mu}^{SM}=288(63)(49)\times 10^{-11}.
\end{equation}
Since the muon AMDM has become a powerful tool to test the validity of the SM and searching for new physics (NP) effects, a plethora of calculations within the framework of  several SM extensions has been reported in the literature in order to explain the $\Delta a_{\mu}$ discrepancy \cite{Jegerlehner:2009ry,Lindner:2016bgg}.

As far as the  $\tau$ lepton is concerned, the SM prediction is $a_{\tau}^{SM}=117721(5)\times 10^{-8}$ \cite{Samuel:1990su,Eidelman:2007sb}. The  error of the order of $10^{-8}$ hints that  SM extensions predicting values for $a_{\tau}$ above this level could be worth studying.   Since the SM  prediction for $a_\tau$ is far from the experimental sensitivity, which is one order of magnitude below  the  leading QED contribution,  a  more precise determination of the experimental value is necessary.  Due to
its short lifetime ($290.3\pm  0.5\times 10^{-15} $ s) \cite{Agashe:2014kda}, the $\tau$ lepton  does not allow for a high precision measurement of its AMDM  via a spin precession method. The most stringent current bound on $a_{\tau}$ is \cite{GonzalezSprinberg:2000mk}
\begin{equation}
-0.052<a_{\tau}<0.013,
\end{equation}
which was obtained using LEP1, SLD and LEP2 data for $\tau$ lepton production. It also has been pointed out recently that the $\tau$ electromagnetic moments can be probed in $\gamma \gamma$ and $\gamma e$ collisions at CLIC, which can lead to improved bounds \cite{Billur:2013rva,Ozguven:2016rst}.   In this regard, it has been pointed out  that super $B$ factories could allow for a precise determination of $a_\tau$ up to the $10^{-6}$ level using unpolarized or polarized electron beams \cite{Bernabeu:2007rr,Fael:2013ij,Eidelman:2016aih}.  Furthermore, due to its large mass, it is expected that the $\tau$ AMDM can be very sensitive to NP effects \cite{Pich:2013lsa} since  its electroweak contribution would be ten times larger than the uncertainty of the hadronic contribution \cite{Eidelman:2007sb}. Therefore, it is worth  estimating the  $\tau$ AMDM in any SM extension as future measurements may allow us to search for NP in a rather clean environment.

 Unlike the electromagnetic dipole moments  of leptons, little attention has been paid to the study of their   weak properties.
 In the experimental arena, the current best limits on the $\tau$  AWMDM and WEDM, with 95\% C.L., are \cite{Heister:2002ik}
 \begin{eqnarray}
Re\left(a_{\tau}^{W}\right) & < & 1.14\times10^{-3},\\
Im\left(a_{\tau}^{W}\right) & < & 2.65\times10^{-3},\\
Re\left(d_{\tau}^{W}\right) & < & 0.5\times10^{-17}\,{\rm ecm},\\
Im\left(a_{\tau}^{W}\right) & < & 1.1\times10^{-17}\,{\rm ecm},
\end{eqnarray}
which were extracted from the data collected at the LEP from 1990 to 1995, corresponding to an integrated luminosity of 155 pb$^{-1}$. Somewhat weaker bounds were obtained in \cite{Hayreter:2013vna}  via a study of the $pp\rightarrow\tau^+\tau^-$ and $pp\rightarrow Zh\rightarrow\tau^+\tau^-h$ cross sections at the LHC.   The current experimental bounds on $a_{\tau}^{W}$ are well above  the SM theoretical prediction, which was calculated in  Ref.
\cite{Bernabeu:1994wh}:

\begin{equation}
 a_\tau^W=-(2.10+0.61i)\times10^{-6}.
\end{equation}
It is thus interesting to analyze whether NP contributions can give a significant enhancement and be at the reach of future experimental detection.

In this work we evaluate the  AMDM and AWMDM of a lepton, with special focus on those of the $\tau$ lepton, predicted by the simplest little Higgs model (SLHM) \cite{Schmaltz:2004de}, which is an appealing SM extension. This model is aimed to  deal with the hierarchy problem by conjecturing that the Higgs boson is a pseudo-Goldstone boson arising from a  global symmetry broken spontaneously. At the same scale,  the local symmetry is also broken by a collective symmetry breaking mechanism. The top quark and the electroweak gauge bosons have heavy partners that give rise to new contributions that exactly cancel the  quadratic divergences to the  Higgs boson mass at the one-loop level, thereby rendering   a mass of about one hundred GeV without the need of fine tuning \cite{ArkaniHamed:2001ca,ArkaniHamed:2001nc}.

The rest of this presentation is organized as follows. A brief review on the SLHM is presented in the Section \ref{model}, whereas  Sec. \ref{analyticalresults} is devoted to the analytical results for  the AWMDM. As a byproduct we will obtain the corresponding expressions for the AMDM. A brief discussion on the current constraints on the parameters of the model, and the  numerical analysis of the $\tau$ electromagnetic and weak dipole moments  is presented in Section \ref{numericalresults}.  Section \ref{conclusions} is devoted to the conclusions, whereas the SLHM Feynman rules as well as  explicit expressions for the loop integrals are shown in the Appendices.
\section{The simplest Little Higgs Model} \label{model}

We now present an overview of the SLHM focusing only on the details relevant for our calculation. For a detailed account of this model and the study of its phenomenology we refer the reader to Refs.  \cite{Schmaltz:2004de,Han:2005ru,delAguila:2011wk,Lami:2016vrs,Lami:2016mjf}, which we will follow closely in our discussion below. The SLHM is the most economic version of  simple-group little Higgs models, which have the feature that the SM gauge group is embedded into a larger simple gauge group instead of a product gauge group. The SLHM has a $\left[SU(3)\times U(1)\right]^{2}$ global symmetry  and a $SU(3)_{L}\times U(1)_{X}$ gauge symmetry, which requires the introduction of nine gauge bosons. At the TeV scale, the global symmetry is broken down spontaneously to $\left[SU(2)\times U(1)\right]^{2}$ via the vacuum expectation values (VEVs) $f_{1}$ and $f_{2}$  of  two sigma fields $\Phi_{1}$ and $\Phi_{2}$,  giving rise to ten Goldstone bosons. At the same scale the gauge group breaks to the SM gauge group $SU(2)_{L}\times U(1)_{Y}$ and five  Goldstone bosons   are eaten by the heavy fields: a charged gauge boson $X^\pm $, a no-self conjugate neutral boson $Y_0\ne Y_0^\dagger$, and an extra neutral gauge boson $Z'$, which thus acquire masses of the order of the scale $f_1\sim f_2$. The remaining Goldstone bosons are accommodated in a complex doublet (the SM one) and a real singlet of $SU(2)$. The Goldstone bosons can be parametrized by the triplets
\begin{equation}
\Phi_{1}=e^{i\Theta\frac{f_{2}}{f_{1}}}\left(\begin{array}{c}
0\\
0\\
f_{1}
\end{array}\right),\;\Phi_{2}=e^{-i\Theta\frac{f_{1}}{f_{2}}}\left(\begin{array}{c}
0\\
0\\
f_{2}
\end{array}\right),
\end{equation}
where the pion matrix is
\begin{equation}
\Theta=\frac{1}{f}\left(\begin{array}{cc}
\begin{array}{cc}
\frac{\eta}{\sqrt{2}} & 0\\
0 & \frac{\eta}{\sqrt{2}}
\end{array} & h\\
h^{\dagger} & \frac{\eta}{\sqrt{2}}
\end{array}\right),
\end{equation}
 here $f=\sqrt{f_1^2+f_2^2}$,  $h$ is the $SU(2)$ complex doublet of the SM, and $\eta$ is a real scalar field. The normalization is chosen to produce canonical kinetic terms. The dynamics of the Goldstone bosons is described by a non-linear sigma model
 \begin{equation}
 \label{Lscakin}
 {\cal L}_{kin} =|D_\mu \Phi_1|^2+|D_\mu \Phi_2|^2,
\end{equation}
with the $SU(3)_L\times U(1)_X$ covariant derivative
\begin{equation}
\label{covderivative}
D_\mu=\partial_\mu+ig T^aA^a_\mu-ig_X Q_X B^{X}_\mu,
\end{equation}
 where $T^a$ ($a=1\dots 8$) are the $SU(3)_L$ generators in the fundamental representation,  $A^a_\mu$ are the  $SU(3)_L$ gauge fields,  $B^{X}_\mu$ is the $U(1)_X$ gauge field, and $Q_X=-1/3$ for $\Phi_1$ and $\Phi_2$.  The new gauge bosons accommodate in a complex $SU(2)\times U(1)$ doublet $(X^{+},\, Y_0)$ with hypercharge $\frac{1}{2}$ and a neutral singlet $Z'_0$.  The matching of the gauge coupling constants yields
\begin{equation}
g_{X}=\frac{gt_{W}}{\sqrt{1-t_{W}^{2}/3}},
\end{equation}
 with $t_{W}=s_{W}/c_{W}$ the tangent of the Weinberg angle $\theta_{W}$.

As mentioned above, after the first stage of symmetry breaking, there emerge the pair of charged gauge bosons $X^\pm $,  the no-self conjugate neutral gauge boson  $Y_0$, and the neutral gauge boson $ Z'_0$ (following  Ref. \cite{Han:2005ru} we will denote the gauge eigenstates with  the subindex  $0$)
\begin{eqnarray}
\label{XField}
X^\pm &=&\frac{1}{\sqrt 2}(A^6\mp iA^7),\\
\label{YField}
Y_0&=&\frac{1}{\sqrt 2}(A^4-iA^5),\\
\label{ZpField}
Z'_0&=&\frac{1}{\sqrt 3}\left(\sqrt{3-t_W^2}A^8 +t_W B^X  \right),
\end{eqnarray}
with masses of the order of $f$. Four gauge fields remain massless at this stage. While  $A^1$, $A^2$, $A^3$ identify with the $SU(2)_L$ gauge bosons $W^a$, the charged gauge bosons $W^\pm $  and the hypercharge gauge boson are
 \begin{eqnarray}
 \label{WField}
W^\pm &=&\frac{1}{\sqrt 2}(A^1\mp i A^2),\\
\label{BField}
B&=&-\frac{1}{3}\left(t_WA^8+\sqrt{3-t_W^2}B^X\right).
\end{eqnarray}

After the electroweak  symmetry breaking (EWSB),  the  weak gauge bosons $W^\pm $ and $Z_0$ acquire mass and the heavy gauge boson masses get corrected. Up to order $(v/f)^2$ $W$,  $X$ and $Y_0$  coincide with the mass eigenstates and their masses are \cite{delAguila:2011wk}

\begin{eqnarray}
m_{W} & = & \frac{gv}{2}\left(1-\frac{v^2}{12f^2}\left(\frac{s_\beta^4}{c_\beta^2}+\frac{c_\beta^4}{s_\beta^2}\right)\right),\\
m_{X} & = & \frac{gf}{\sqrt{2}}\left(1-\frac{v^{2}}{4f^{2}}\right),\\
m_Y&=& \frac{gf}{\sqrt{2}},
\end{eqnarray}
with $t_\beta=\tan\beta=f_1/f_2$. If higher order terms are considered $W$ and $X$ need to be rotated to obtain the physical states \cite{delAguila:2011wk}. On the other hand, the photon and the light  neutral $Z_0$ gauge boson  are given by
\begin{eqnarray}
\label{AField}
A &=& s_W A^3 +c_W B,\\
\label{ZField}
Z _0&=&c_W A^3  -s_W B.
\end{eqnarray}
Finally, $Z_0$ and $Z'_0$ need to be rotated to obtain the mass eigenstates $Z$ and $Z'$, which are given by
 \begin{eqnarray}
 \label{Zpeigen}
Z' _0& =& Z'+\delta_{Z}Z,\\
\label{Zeigen}
Z_0 & =& Z-\delta_{Z}Z',
\end{eqnarray}
with  $\delta_{Z}=-\frac{(1-t_{W}^{2})\sqrt{3-t_{W}^{2}}}{8c_{W}}\frac{v^{2}}{f^{2}}$. The respective masses, up to order $(v/f)^2$, are \cite{delAguila:2011wk}
\begin{eqnarray}
m_{Z} & = & \frac{gv}{2c_{W}}\Bigg(1-\frac{v^2}{12f^2}\left(\frac{s_\beta^4}{c_\beta^2}+\frac{c_\beta^4}{s_\beta^2}\right)\nonumber\\
&-&\frac{v^2}{16f^2}\left(1-t_W^2\right)^2\Bigg),\\
m_{Z'} & = & \frac{\sqrt{2} gf}{\sqrt{3-t_{W}^{2}}}\left(1-\frac{v^{2}}{f^{2}}\frac{(3-t_{W}^{2})}{16c_{W}^{2}}\right).
\end{eqnarray}

The kinetic Lagrangian of the gauge bosons  gives rise to the trilinear gauge boson couplings  necessary for our calculation. It can be written as
\begin{equation}
{\cal L}^G=-\frac{1}{4}{B^X}^{\mu\nu}B^X_{\mu\nu}-\frac{1}{4}{A^a}^{\mu\nu}{A^a}_{\mu\nu},
\end{equation}
with the Abelian and non-Abelian gauge strength  tensors
\begin{equation}
B^X_{\mu\nu}=\partial_\mu B^X_\nu-\partial_\nu B^X_\mu,
\end{equation}
and
\begin{equation}
A^a_{\mu\nu}=\partial_\mu  A^a_\nu-\partial_\nu  A^a_\mu+gf_{abc}A^b_\mu A^c_\nu,
\end{equation}
with $f_{abc}$ the structure constants of the $SU(3)$ group. From the  relations between gauge eigenstates  and mass eigenstates (\ref{XField})-(\ref{BField}) and (\ref{AField})-(\ref{Zeigen}) we can obtain after some lengthy algebra the Feynman rules listed in  \ref{FeynmanRules} for the $VV_j^\pm V_j^\mp$ vertices, namely $AW^\pm W^\mp$, $AX^\pm  X^\mp$, $ZW^\pm W^\mp$, and $ZX^\pm X^\mp$.

In the lepton sector of the SLHM, for each generation there is a left-handed triplet $L_m^T=(\nu_{L_m},\ell_{L_m}, iN_{Lm})$,  which is completed with a new neutral lepton $N_{Lm}$, and two right-handed singlets $\ell_{Rm}$ and $N_{Rm}$. The Yukawa Lagrangian can be written, in the basis where flavor and mass $N_m$ eigenstates   coincide, as
\begin{equation}
\label{LYukawa}
\mathcal{L}_{Y}= i\lambda_{N}^{m}\overline{N}_{Rm}\Phi_{2}^{\dagger}L_{m}+\frac{i\lambda_{\ell}^{mn}}{\Lambda}\overline{\ell}_{Rm}\epsilon_{ijk}\Phi_{1}^{i}\Phi_{2}^{j}L_{n}^{k}
+{\rm H.c.},
\end{equation}
where $\Lambda = 4\pi f$ is the cut-off of the effective theory. Here $m$ and  $n$ are generation indices, whereas $i$, $ j$, and $ k$ are $SU(3)$ indices. After EWSB this Lagrangian yields the lepton ma\-sses and the heavy neutrino masses up to order $(v/f)^2$ \cite{delAguila:2011wk}

\begin{eqnarray}
\mathcal{L}_{mass} & =& -fs_{\beta}\lambda_{N}^{m}\Bigg(\left(1-\frac{\delta_{\nu}^{2}}{2}\right)\overline{N}_{Rm}N_{Lm}\nonumber\\
&-&\delta_{\nu}\overline{N}_{Rm}\nu_{Lm}\Bigg)\nonumber \\& + & \left(1-\frac{v^{2}}{4f^{2}}-\frac{v^{2}}{12f^{2}}\left(\frac{s_{\beta}^{4}}{c_{\beta}^{2}}+\frac{c_{\beta}^{4}}{s_{\beta}^{2}}\right)\right)
 \frac{fv}{\sqrt{2}\Lambda}\nonumber\\&\times&\lambda_{\ell}^{mn}\overline{\ell}_{Rm}\ell_{Ln}+{\rm H.c.},
\end{eqnarray}
where  $ \delta_{\nu}=\frac{v}{\sqrt{2}ft_{\beta}}$ represents the mixing between a heavy neutrino and a SM neutrino of the same generation.   Notice also that the rotation that diagonalizes $\lambda_N$ does not necessarily diagonalizes $\lambda_\ell$ so there is mixing between the charged leptons and the heavy neutrinos mediated by the charged gauge bosons.
The charged lepton mass eigenstates $\ell_{Lm}$ are thus related to the flavor eigenstates $\ell_{Lm0}$ by the rotation
\begin{equation}
\ell_{Lm0} = V^{mi}\ell_{Li} .
\end{equation}
where $V^{mi}$ is a CKM-like mixing matrix. Also, in each generation, the SM and heavy neutrino mass eigenstates are obtained through
\begin{eqnarray}
\nu_{Li0}&=& \left(1-\frac{\delta_\nu^2}{2}\right)\nu_{Li}-\delta_\nu V^{im} N_{Lm},\\
N_{Lm0}&=&\left(1-\frac{\delta_\nu^2}{2}\right) N_{Lm}+  \delta_\nu V^{mi}\nu_{Li},
\end{eqnarray}
where again the $0$ subindex stands for flavor eigenstates.
The lepton masses up to order $(v/f)^2$ are \cite{delAguila:2011wk}
\begin{equation}
m_{\ell}=-
\left(1-\frac{v^{2}}{4f^{2}}-\frac{v^{2}}{12f^{2}}\left(\frac{s_{\beta}^{4}}{c_{\beta}^{2}}+\frac{c_{\beta}^{4}}{s_{\beta}^{2}}\right)\right)\frac{fv}{\sqrt{2}\Lambda}y_{\ell},
\end{equation}
where $y_{\ell}$ is the eigenvalue of the $\lambda_\ell$ matrix and
\begin{equation}
m_{N_{i}}=fs_{\beta}\lambda_{N}^{i}.
\end{equation}

The vertices of a gauge boson to a lepton pair are obtained from the lepton kinetic Lagrangian, which can be written as
\begin{eqnarray}
{\cal L}_F&=&\bar{L}_m i\gamma^\mu D_\mu L_m+\bar{\ell}_{Rm} i\gamma^\mu D_\mu \ell_{Rm}\nonumber\\&+&\bar{N}_{Rm} i\gamma^\mu N_\mu N_{Rm},
\end{eqnarray}
where the covariant derivative was given in Eq. (\ref{covderivative}), with $Q_X=-1/3$, $0$ and $1$ for $L_m$, $N_m$ and $\ell_m$. We need to introduce the mass eigenstates to obtain the  interactions of the physical gauge bosons $Z$, $W$, $X$, and $Z'$ to a  lepton pair, which are necessary for our calculation of the AMDM and AWMDM of a lepton. They  are given by

\begin{eqnarray}
{\cal L}_{Zff'}&=&-\frac{g}{c_W} Z_\mu \Bigg[
\left(-\frac{1}{2}+s_W^2\right)\bar{\ell}_{Li}\gamma^\mu \ell_{Li}\nonumber\\&+&\frac{1}{2}\left(1-\delta_\nu^2\right)\bar{\nu}_{Li}\gamma^\mu \nu_{Li}+s_W^2 \bar{\ell}_{Ri}\gamma^\mu\ell_{Ri}
 \nonumber\\&+&\frac{1}{2}\delta_\nu^2\bar{N}_{Li}\gamma^\mu N_{Li}-\frac{1}{2}\left(\delta_\nu V^{im}\bar{N}_{Lm}\gamma^\mu\nu_{Li}+{\rm H.c.}\right)\nonumber\\&+&\frac{\delta_Z}{c_W\sqrt{3-t_W^2}}\Bigg(\left(\frac{1}{2}-s_W^2\right)
 \left(\bar{\ell}_{Li}\gamma^\mu\ell_{Li}+\bar{\nu}_{Li}\gamma^\mu\nu_{Li}\right)
\nonumber\\&-&s_W^2 \bar{\ell}_{Ri}\gamma^\mu\ell_{Ri}-c_W^2\bar{N}_{Li}\gamma^\mu N_{Li}\Bigg)\Bigg],\nonumber\\
\end{eqnarray}

\begin{eqnarray}
{\cal L}_{Wf f'}&=&-\frac{g}{\sqrt 2} W_\mu^-\left( \left(1-\frac{\delta_\nu^2}{2}\right)\bar{\ell}_{Li}\gamma^\mu \nu_{Li}\right.\nonumber\\
&-&\left.\frac{1}{2}\delta_\nu V^{im}\bar{\ell}_{Li}\gamma^\mu  N_{Lm}\right)+{\rm H.c.},
\end{eqnarray}

\begin{eqnarray}
{\cal L}_{Xf f'}&=&-i\frac{g}{\sqrt 2} X_\mu^- \Bigg( \delta_\nu \bar{\ell}_{Li}\gamma^\mu\nu_{Li}\nonumber\\&+&\left(1-\frac{\delta_\nu^2}{2}\right)\bar{\ell}_{Li} \gamma^\mu V^{im} N_{Lm}+\Bigg)+{\rm H.c.} ,
\end{eqnarray}
and
\begin{eqnarray}
{\cal L}_{Z\ell \ell}&=&\frac{g}{c_W^2\sqrt{3-t_W^2}} Z'_\mu \Bigg[\left(-\frac{1}{2}+s_W^2\right)\bar{\ell}_{Li}\gamma^\mu \ell_{Li}\nonumber\\&+&s_W^2\bar{\ell}_{Ri}\gamma^\mu\ell_{Ri}\nonumber\\&+&\delta_Z c_W\sqrt{3-t_W^2}\Bigg(\left(-\frac{1}{2}+s_W^2\right)\bar{\ell}_{Li}\gamma^\mu\ell_{Li}
\nonumber\\&+&s_W^2 \bar{\ell}_{Ri}\gamma^\mu\ell_{Ri}\Bigg)\Bigg].
\end{eqnarray}
Notice that there is lepton flavor violation mediated by the charged gauge bosons. Finally, the interactions of the photon with a charged lepton pair are dictated by QED

\begin{equation}
{\cal L}_{A\ell \ell}=-e A_\mu \bar{\ell}\gamma^\mu\ell.
\end{equation}

The scalar Higgs bosons $H$ and $\eta$ also contribute to a lepton AMDM and  AWMDM. From the Lagrangian (\ref{Lscakin}) we can  obtain the respective  interactions with  the $Z$ and $Z'$ gauge bosons. After some algebra one can extract the  vertices  $ZZ'H$ and $ZZH$, which are given as follows to the leading order in $(v/f)$ as

\begin{equation}
{\cal L}_{ZZ'H}=\frac{g^2v(1-t_W^2)}{2c_W\sqrt{3-t_W^2}} H Z^\mu Z'_\mu,
\end{equation}
together with the $ZZH$ interaction
\begin{equation}
{\cal L}_{ZZH}=gm_Z\left(1-\frac{v^2}{4f^2}\left(\frac{1}{3}
\left(\frac{s_\beta^4}{c_\beta^2}+\frac{c_\beta^4}{s_\beta^2}\right)\right)\right)
H Z^\mu Z_\mu.
\end{equation}
From the Yukawa Lagrangian (\ref{LYukawa}) we can also obtain the interactions of the scalar Higgs bosons $H$ and  $\eta$ to leptons, which we need for our calculation. They are diagonal and are given to leading order in $(v/f)^2$ by \cite{Han:2005ru}
\begin{equation}
{\cal L}_{H\bar{\ell}\ell}=ig\frac{m_{\ell}}{2m_W}  \left(1-\frac{v^2}{6f^2}\left(\frac{1}{t_\beta}+t_\beta\right)^2\right)\bar{\ell} \ell  H,
\end{equation}
and
\begin{equation}
{\cal L}_{\eta\bar{\ell}\ell}=ig\frac{m_{\ell}}{\sqrt{2}f} \left(\frac{1}{t_\beta}-t_\beta\right) \bar{\ell}\gamma^5\ell \eta.
\end{equation}

The $\eta ZZ'$ vertex vanish since  $\eta$ is $CP$-odd scalar,  but the $H\eta Z$ coupling does arise, though its contribution to the AMDM and AWMDM of a lepton vanishes. A similar result was found in Ref. \cite{Bernabeu:1995gs}, where the contributions of two-Higgs doublet models (THDMs) to the AWMDM of a fer\-mion were calculated.

Other details of this model are  irrelevant for our calculation, so we refrain from presenting a discussion of the quark sector and the Coleman-Weinberg scalar potential.


\section{Anomalous magnetic and weak  magnetic dipole moments in the SLHM}
\label{analyticalresults}
We now turn to present our results. All the Feynman rules necessary for our calculation  follow straightforwardly from the above interaction Lagrangians and are presented in  \ref{FeynmanRules}. Since we are interested in the $\bar{\ell}\ell V^\mu$ vertex with all the particles on their mass shell, the loop amplitudes will be gauge independent.  We used the unitary gauge  as it is best suited for our calculation method. In order to solve the loop integrals, we  used both Feynman parametrization and the Passarino-Veltman reduction scheme. We will first present the results for the  AWMDM, from which the results for the AMDM will follow easily. As far as the $CP$-violating properties are concerned,  we will assume that there is no new sources of $CP$-violation in this model's version, so  both the EDM and  the WEDM will vanish. In the most general scenario, $CP$-violation could arise from new additional phases in the extended Yukawa sector of the SLHM.

\subsection{Anomalous weak magnetic dipole moment }

\begin{figure}[hbt!]
\begin{center}
\includegraphics[width=8.25cm]{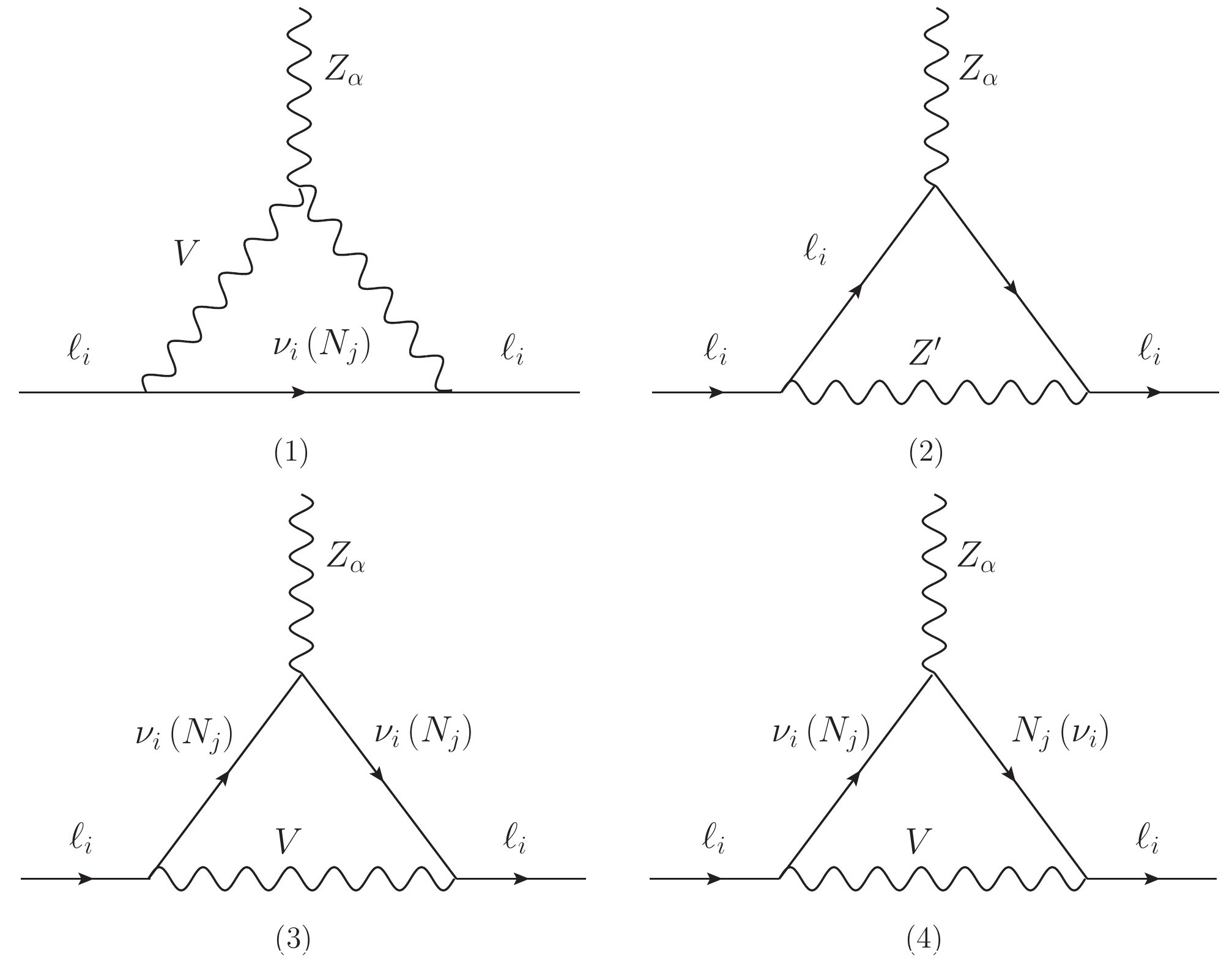}
\end{center}
\caption{Feynman diagrams that contribute to   the AWMDM of  charged lepton $\ell_i$ at the one-loop level  in the gauge sector of the SLHM. Here $V$ can be either the $W$ gauge boson or the new charged $X$ gauge boson, $\nu_i$  and $N_j$ stand for a  SM neutrino and a new heavy one predicted by the SLHM, respectively. Notice that diagram (4) involves the  non-diagonal vertex $Z \bar{\nu}_i N_j$.  \label{WMDMGDiagrams}}
\end{figure}

In the SLHM, in addition to the pure SM contributions, the AWMDM receives new physics contributions arising from the loops carrying only new particles, but also from loops involving only SM particles. The latter are due to corrections to the SM vertices and appear as a series of powers of $v/f$, so it is enough to consider the leading order terms. The AWMDM of lepton $\ell_i$ can thus by written as

\begin{equation}
\label{aWSLHM}
a^{W-SLHM}_{\ell_i}=a^{W-SM}_{\ell_i}+ a^{W-NP}_{\ell_i},
\end{equation}
where $a^{W-SM}_{\ell_i}$ stands for the SM contributions and $a^{W-NP}_{\ell_i}$  for the new physics ones, which can be written as

\begin{equation}
a^{W-NP}_{\ell_i}=a^{W-Gauge}_{\ell_i}+a^{W-Scalar}_{\ell_i},
\label{aWNP}
\end{equation}
with $a^{W-Gauge}_{\ell_i}$ $\left(a^{W-Scalar}_{\ell_i}\right)$ the contributions arising from the gauge (scalar) sector of the SLHM.

In the gauge sector, the NP contributions to the AWMDM arise from the Feynman  diagrams of Fig. \ref{WMDMGDiagrams}, where    $V$ stands for the charged  gauge bosons $W$ and $X$,  $\nu_i$ is a SM neutrino, and $N_k$ is a heavy neutrino.   According to our discussion above, for the loops involving only  $W$ gauge bosons and  SM neutrinos we consider  the leading order $v/f$ contributions arising from the corrections to the $W\bar\ell \nu $ vertex. The  contributions to the AWMDM of each Feynman diagram will be written as follows
\begin{equation}
\label{a^W_ell}
a_{\ell_i}^{W-A_1A_2A_3}=\frac{\alpha }{4\pi}f^{A_1A_2A_3}_ZI^{A_1 A_2 A_3}_Z
\end{equation}
where  the three-letter superscript   stands for the particles circulating  in each loop diagram ($A_2$ and $A_3$ are the particles attached to the external $Z$ gauge boson, whereas $A_1$ is the particle attached to the external leptons). Here  $f^{A_1A_2A_3}_Z$ are  coefficients involving all the couplings appearing in each  amplitude, whereas  $I^{A_1A_2A_3}_Z$ stand for the loop integrals, which depend on the masses of the virtual particles. We  present in  \ref{LoopIntegrals} both  the $f^{A_1A_2A_3}_Z$ coefficients and the  loop integrals  in terms of parametric integrals and Passarino-Veltman scalar functions.  We have verified that the contribution of each diagram to the AWMDM is free of ultraviolet divergences.  The full contribution of the gauge sector can be written as
\begin{eqnarray}
a_{\ell_i}^{W-Gauge}&=&\sum_{V=W,X}\sum_{n=\nu_i,N_j}\left(a_{\ell_i}^{W-n VV}+a_{\ell_i}^{W-V n n}\right)\nonumber\\&+&a_{\ell_i}^{W-Z'\ell_i\ell_i},
\end{eqnarray}
where the index $j$  runs over the three lepton families.

\begin{figure}[hbt!]
\begin{center}
\includegraphics[width=8.25cm]{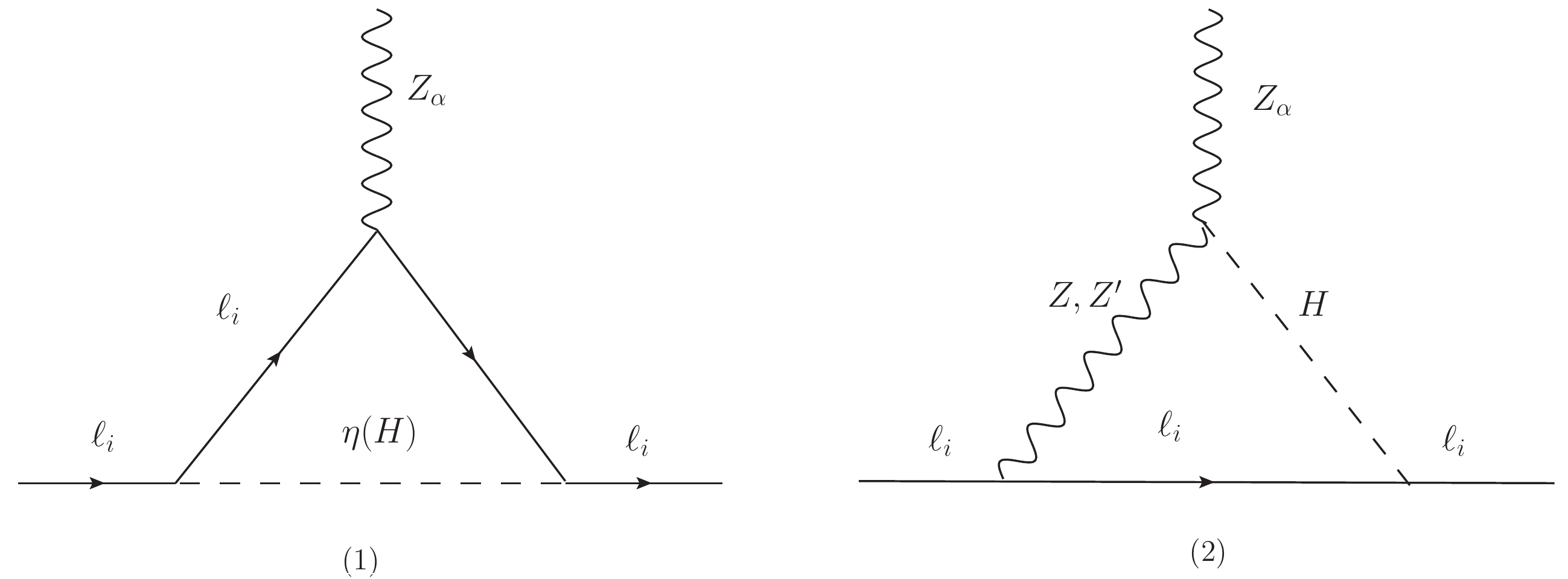}
\end{center}
\caption{Feynman diagrams that contributes to the AWMDM of a lepton in the scalar sector of the  SLHM at the one-loop level.  We do not show the diagram obtained by exchanging the $Z (Z')$ gauge boson and the  Higgs boson in diagram (2).  The new contributions of the SM Higgs boson contributions are due to the diagram with the $Z'$ gauge boson  and also to corrections  to the SM vertices $H\ell\ell$ and  $HZZ$. \label{WMDMSDiagrams}}
\end{figure}

In  the scalar sector of the SLHM there are contributions to the AWMDM of a lepton arising from both the SM scalar boson $H$ and  the new pseudoscalar boson $\eta$  via the Feynman diagrams of Fig. \ref{WMDMSDiagrams}.   The contributions of the SM Higgs boson  arise from corrections of the order of $(v/f)^2$  to the SM vertices $H\ell\ell$ and  $HZZ$. The respective contributions to the AWMDM can also be written as in Eq. (\ref{a^W_ell}), where the $f^{A_1A_2A_3}_Z$ coefficients and the $I^{A_1A_2A_3}_Z$  functions are presented in  \ref{LoopIntegrals}.  Again we have verified that the contribution of each diagram to the AWMDM is  ultraviolet finite.

The full scalar contribution is thus
\begin{eqnarray}
a_{\ell_i}^{W-Scalar}&=&\sum_{S=\eta,H}a_{\ell_i}^{W-S\ell_i\ell_i }+\sum_{V=Z,Z'}a_{\ell_i}^{W-\ell_i HV}.
\end{eqnarray}

\subsection{Anomalous magnetic dipole moment}
In the gauge sector of the SLHM, the AMDM of a lepton arises from the Feynman diagrams (1) and (2) of Fig. \ref{WMDMGDiagrams}, with the $Z$ gauge boson replaced by the photon. There are also contributions  arising  from the scalar sector, which are induced by the scalar bosons $H$ and $\eta$ via a Feynman diagram similar to diagram (1) of  Fig. \ref{WMDMSDiagrams}. The corresponding contributions to the AMDM can be obtained straightforwardly from those to the AWMDM by considering the limit $m_Z\to 0$ and substituting the $Z$ coupling constants by those of the photon.  We can write  the contributions to $a_\ell$ arising from each diagram as
\begin{equation}
\label{a_ell}
a_{\ell_i}^{A_1A_2A_3}=\frac{\alpha f^{A_1A_2A_3}_\gamma}{4\pi} I^{A_1A_2A_3}_\gamma,
\end{equation}
where again the  three-letter superscript corresponds to  the three particles circulating in the loop. Explicit expressions for the  $f^{A_1A_2A_3}_\gamma$ constants and the $I^{A_1A_2A_3}_\gamma$ functions are presented in  \ref{LoopIntegrals} in terms of parametric integral and Passarino-Veltman scalar functions.

The overall NP contribution of the SLHM to $a_{\ell_i}$ is thus
\begin{eqnarray}
a_{\ell_i}^{NP}&=&  \sum_{V=W,X}\sum_{n=\nu_i,N_j}a_{\ell_i}^{n VV}+a_{\ell}^{Z' \ell_i\ell_i}+\sum_{S=H,\eta}a_{\ell_i}^{S\ell_i \ell_i },
\end{eqnarray}
where the index $j$ runs over the three lepton families.

It is worth mentioning that our results for the $\tau$ AMDM in terms of parametric integrals,  obtained by a limiting procedure from our results for the AWMDM,  agree with previous calculations  presented in the literature  \cite{Leveille:1977rc,Jegerlehner:2009ry}. This serves as a cross-check for our calculation.

\section{Numerical analysis}
\label{numericalresults}

\begin{table*}[!ht]
\begin{center}
\caption{Bound on the symmetry breaking scale $f$ of the SLHM obtained from several observables.\label{boundonf}}
\begin{tabular}{cc}
\hline
\hline
Observable&Lower bound on $f$ (TeV)\\
\hline
\hline
Parity violation in Cesium&$1{.}7$  \cite{Schmaltz:2004de}\\
LEP-II data&$2$  \cite{Schmaltz:2004de}\\
 $Z'$ corrections to the oblique $S$ parameter&$5{.}2$  (95\% C.L.) \cite{Marandella:2005wd}\\
Electroweak precision data&$5{.}6$  (95\% C.L.) \cite{Dias:2007pv}\\
\hline
\hline
\end{tabular}
\end{center}
\end{table*}

We now present our numerical results for the AMDM and  AWMDM of the $\tau$ lepton in the context of the SLHM.
We will  briefly  review the existing bounds on the free parameters of the model and afterwards analyze the potential contributions to the AMDM and the AWMDM of the $\tau$ lepton for   parameter values consistent with these bounds.

\subsection{Bounds on the parameter space of the SLHM}
The SLHM parameters involved in our calculation are  $f$, $t_\beta$, $m_{N_k}$, $m_{\eta}$, $\delta_\nu$, and the matrix elements $V_l^{mi}$. We will   discuss the current bounds on these parameters obtained from the study of experimental data  of several observables as reported in the literature.

{\it  Symmetry breaking scale} $f$:  bounds on this parameter arise  from several observables. We list the most relevant in Table \ref{boundonf}. We can observe that the most stringent bound $f\geq 5.6$ TeV  arises from electroweak precision data (EWPD), whereas the weakest limit $f\geq 1.7$ TeV arises from parity violation in Cesium.

{\it $f_1$ to $f_2$ ratio} $t_\beta$: a fit on 21 electroweak precision observables from LEP, SLC, Tevatron, and the Higgs boson data reported by the LHC collaborations ATLAS and CMS, allowed the authors  of  Ref. \cite{Reuter:2012sd,Reuter:2013zja} to find  out the allowed region in the $t_\beta$-$f$ plane, which we will take into account for our numerical analysis. For the strongest bound $f\geq$ 5{.}6 TeV, the allowed interval of $t_\beta$ values is 1-9.  We will analyze below the dependence on $t_\beta$ of the $\tau$ AMDM and AWMDM  in the allowed interval.

{\it Mixing   between light and heavy neutrinos} $\delta_{\nu}$: this parameter
is experi\-men\-tal\-ly constrained to be small \cite{delAguila:2011wk}, with the corresponding  bound being flavor dependent: $\delta_{\nu_{e}}\leq 0.03$, $\delta_{\nu_{\mu}}\leq 0.05$, and  $\delta_{\nu_{\tau}}\leq 0.09$ with  95\% C.L. Since we are interested in the study of the $\tau$ lepton,  we need to make sure that we use values of $f$ and $t_\beta$ consistent with the bound  $\delta_{\nu}=v/(\sqrt{2}ft_\beta)\leq 0.09$, which in turn translates into the bound $ft_\beta\gtrsim 1932$ GeV. Such a constraint is fulfilled for the  values of $f$ and $t_\beta$ chosen in our analysis.

{\it Pseudoscalar mass} $m_\eta$: this parameter is basically dependent on the $\mu$ parameter ($m_{\eta}\sim \mu$) appearing in the scalar potential via the term $-\mu^2 (\Phi_1^{\dagger}\Phi_2 +H.c.)$. In our analysis we will explore values consistent with the lower bound $m_\eta \leq 7$ GeV, which arises from  the non-observation of the $\Upsilon\to\eta\gamma$ decay \cite{Balest:1994ch}. Although the dominant contribution to the AMDM can arise from a very light pseudoscalar, with mass of the order of 10 GeV, this requires relatively large values of $t_{\beta}$, which are already excluded according to the above discussion \cite{Reuter:2012sd,Reuter:2013zja}.

{\it Mixing matrix elements} $V_\ell^{mi}$: previous studies on LFV  within the SLHM \cite{delAguila:2011wk,Lami:2016vrs,Wang:2012zza,Han:2011aq,Lami:2016mjf} have parametrized the mixing matrix $V_\ell$ and considered bounds on the mixing angles from experimental data on LFV processes. A simple approach was taken by the authors of Ref. \cite{delAguila:2011wk} in which an scenario with mixing between the first and second families only was considered. It was found that the respective angle is  tightly constrained: the limit on  $e-\mu$ conversion yields an upper bound on $\sin 2\theta_{12}$ of the order of $0.005$. As we are interested in the $\tau$ AMDM and AWMDM, we will assume an scenario with mixing between the second and third families only, namely we will consider the following mixing matrix

\begin{equation}
\label{Mixing-Matrix}
V_\ell\simeq \left(
\begin{array}{ccc}
1 & 0 & 0 \\
0&\cos\theta & \sin\theta \\
0 & -\sin\theta & \cos\theta
\end{array}\right).
\end{equation}

\begin{figure}[!hbt]
\centering
\includegraphics[width=8.25cm]{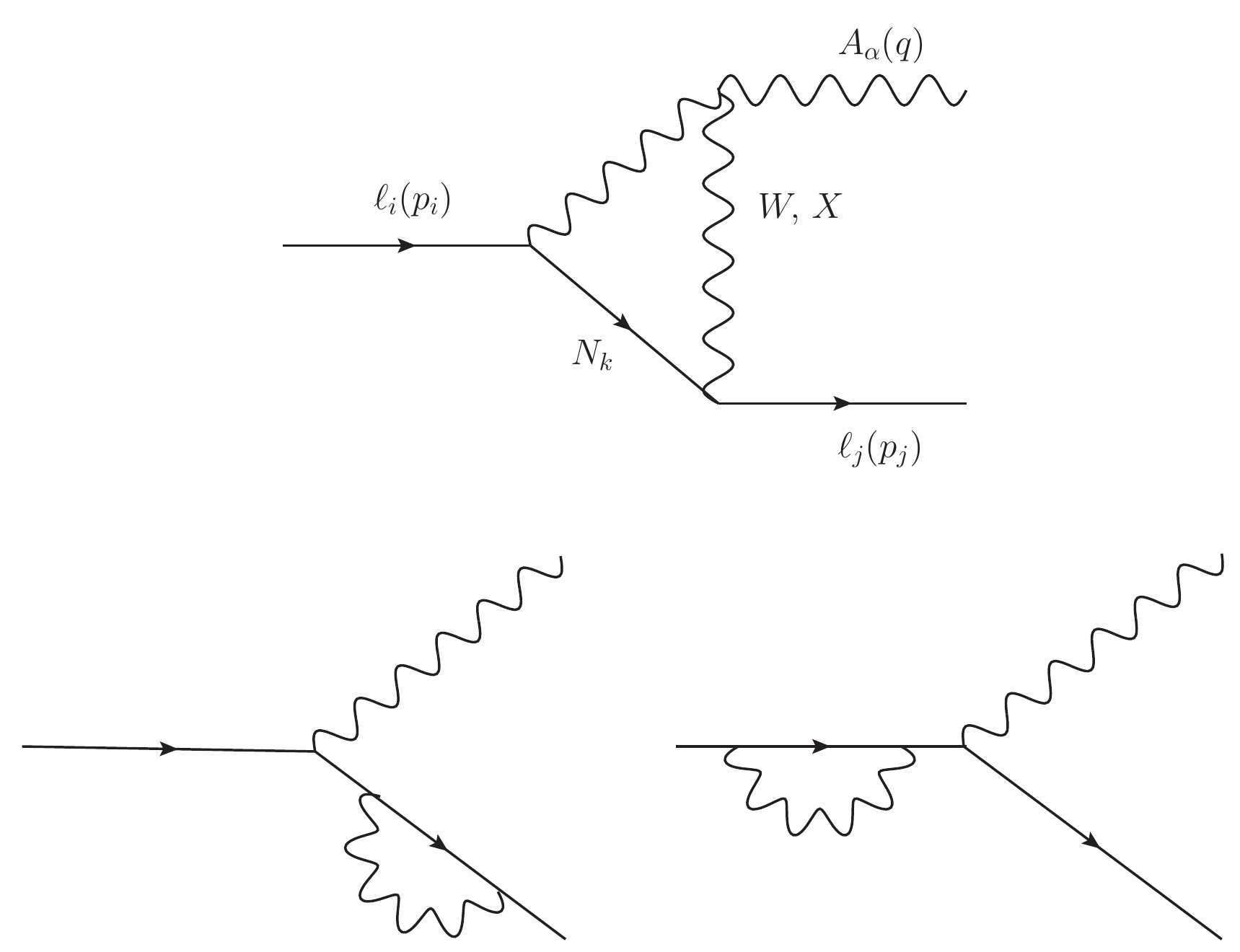}
\caption{SLHM contribution to the $\ell_i\to \ell_j\gamma$ decay at the one-loop level. In the loop circulate the charged $W$ and $X$ gauge bosons and a heavy neutrino $N_k$.
\label{litoljgammadiagram}}
\end{figure}

\begin{figure}[!hbt]
\centering
\includegraphics[width=8.25cm]{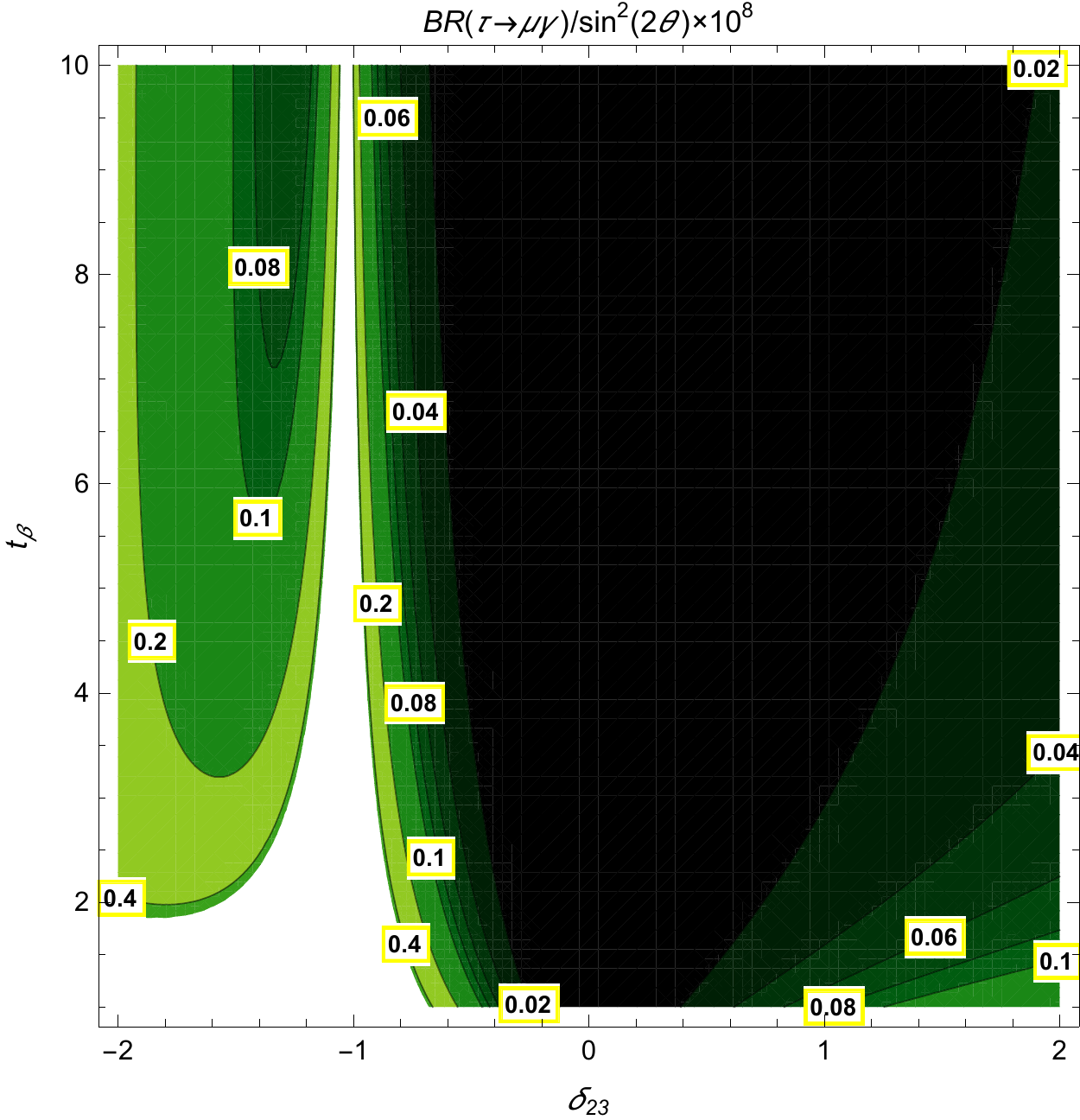}
\caption{Contours of the SLHM contribution to the  $\tau\to \mu\gamma$ branching ratio in the $z_3$ vs
$t_\beta$ plane. We set $z_2=0.5$ and $f=2000$ GeV.
\label{tautomugammaBR}}
\end{figure}

We will analyze whether current experimental bounds on processes such as the muon AMDM  and the  $\tau \to \mu\gamma$ decay can be helpful to find a bound on the mixing angle $\theta$. We already have presented the results for the AMDM and AWMDM of a lepton, we will now present the  SLHM contribution to the  $\tau \to \mu\gamma$ decay in terms of both parametric integrals and Passarino-Velt\-man scalar functions. The Feynman diagrams inducing this decay at the one-loop level  are shown in Fig. \ref{litoljgammadiagram}. The corresponding amplitude  can be written, in the limit of massless $\ell_j$, as

\begin{equation}
\label{litoljgammaamplitude}
{\cal M}(\ell_i\to \ell_j \gamma)=F_L\bar{u}(p_i)\left(\frac{i}{2m_{\ell_i}}\sigma^{\mu\nu}q_\nu\right)P_L u(p_j),
\end{equation}
with $q=p_i-p_j$ and $F_L$  given by

\begin{eqnarray}
F_L&=&\frac{e\alpha}{8 \pi s_W^2}\sum_{V=W,X}\sum_k \delta_{NNV}^2 V_\ell^{*kj}V_\ell^{k i} f_L(x_k)
\nonumber\\&=&\frac{e\alpha}{16 \pi s_W^2}\sin2\theta\sum_{V=W,X}\delta_{NNV}^2\left(f_L(x_{\mu})- f_L(x_{\tau})\right),\nonumber\\
\label{FL}
\end{eqnarray}
where $x_k=(m_{N_k}/m_V)^2$ ($V=W,X$), $\delta_{NNV}=\delta_{\nu}$ for the $W$ gauge boson, and  $\delta_{NN V}=\left(1-\frac{\delta^2_{\nu}}{2}\right)$  for the $X$ gauge boson. The $f(x_{k})$ function is presented in \ref{litoljgammaappendix}. We have verified that our results are in agreement with \cite{delAguila:2011wk}.  The $\ell_i\to \ell_j \gamma$ decay width is given by

\begin{equation}
\label{litoljgammadecaywidth}
\Gamma(\ell_i\to \ell_j \gamma)=\frac{m_{\ell_i}|F_L|^2}{32\pi}.
\end{equation}
The current experimental limit ${\rm BR}(\tau \to\mu\gamma)\leq  4.4 \times 10^{-8}$ \cite{Olive:2016xmw} can translate into a  bound on $\sin2\theta$. For this purpose, we introduce the mass splitting $\delta_{23}\equiv z_3-z_2$ with $z_k=(m_{N_k}/m_X)^2$ and show in  Fig. \ref{tautomugammaBR} the contours of the branching ratio of the $\tau\to \mu\gamma$ decay in the $\delta_{23}$ vs $t_\beta$ plane for $z_2=1$ and $f=2000$ GeV. We conclude that even for a large splitting $\delta_{23}$, the branching ratio ${\rm BR}(\tau \to\mu\gamma)$ would hardly reach a level above $10^{-8}$  for $\sin 2\theta$ of the order of unity, thereby yielding a very weak constraint on this parameter. On the other hand, the muon AMDM also does not  yields a useful bound on $\theta$ as the heavy neutrino contribution is negative and cannot account for the muon AMDM discrepancy of Eq. (\ref{discrepancy}).

As we will see below, the only relevant contribution to the AMDM and WAMDM involving the mixing matrix elements is the contribution of the heavy neutrino, which has the generic form

\begin{eqnarray}
a_\tau^{N_k XX}&\sim&\sum_{V=W,X}\sum_i V_\ell^{*i3}V_\ell^{i 3} F(x_{k})\nonumber\\&=&
\sum_{V=W,X}\left(\cos\theta^2 F(x_{\tau})+\sin^2\theta F(x_{\mu})\right)\nonumber\\&=&
\sum_{V=W,X}\left( F(x_{\tau})+\sin^2\theta \left(F(x_{\mu})-F(x_{\tau})\right)\right).
\end{eqnarray}
It turn out that the second term is subdominant for a small mixing angle $\theta$ and a small splitting $\delta_{23}$.  Following the authors of Ref. \cite{Lami:2016vrs} in their study of LFV hadronic $\tau$ decays, we will consider values of $\sin 2\theta$  of the order of  $10^{-1}$. Under this assumption,   the  term proportional to $\sin^2\theta$  becomes subdominant, which is equivalent to consider an approximately diagonal mixing matrix. In addition, we have found out that there is little sensitivity of our results to a small change in the value of $\sin2\theta$. A larger value of this parameter would increase slightly  the $\tau$ AMDM and AWMDM, but an enhancement larger than one order of magnitude would hardly be attained.

\begin{figure}[!hbt]
\centering
\includegraphics[width=8.25cm]{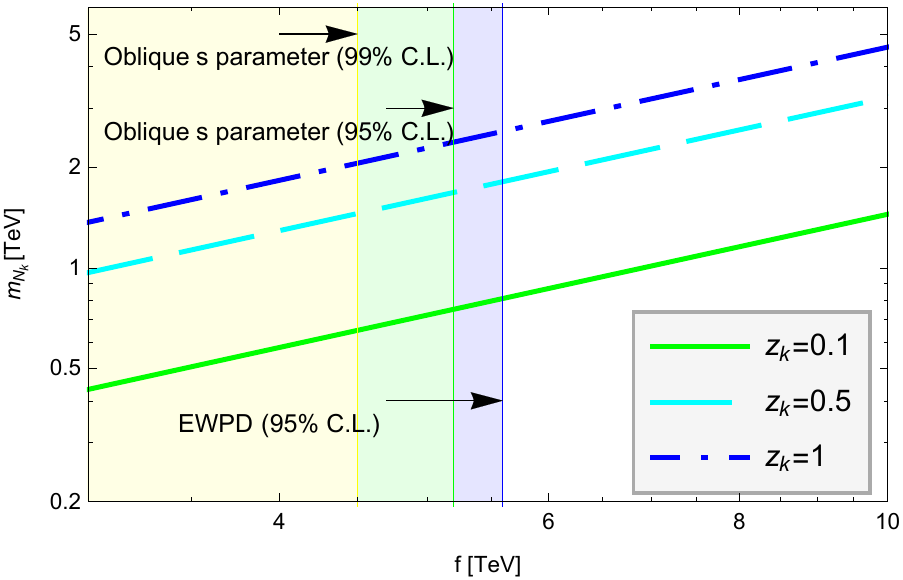}
\caption{Heavy neutrino mass as a function of $f$ for  $t_\beta=9$ and  three values of $z_k=(m_{N_k}/m_X)^2$.   The vertical lines represent the lower bounds on $f$ arising from several observables (see Table \ref{boundonf}).
\label{Nmass}}
\end{figure}

{\it Heavy neutrino mass}: we will follow  the approach of Ref. \cite{delAguila:2011wk}, in which   $m_{N_k}$ is parametrized through the ratio  $z_k=(m_{N_k}/m_X)^2$. As observed in Fig. \ref{Nmass}, for the most stringent limit  $f=5.6$ TeV,   $m_{N_k}\geq 0.836$, which corresponds to the value $z_k=0.1$, whereas $m_{N_k}\geq 2.644$ TeV  for $z_k=1$.  Again, the $\tau$ AMDM and AWMDM contributions arising from the heavy neutrinos show little sensitivity to  moderate changes in the value of $z_3$ and the mass splitting $\delta_{23}$, so we will use as reference values $z_3\simeq 1$ and $\delta_{23}\leq 0.5$.

In conclusion, we will use the set of values shown in Table \ref{values} for the SLHM  parameters involved in our numerical analysis. We have found that except for $f$ and $t_\beta$ there is little sensitivity  of the $\tau$ AMDM and AWMDM    to a mild change in the values of the remaining parameters as far as they lie between the allowed intervals.

\begin{table}[!ht]
\begin{center}
\caption{Values used in our numerical analysis for the parameters involved in the AMDM and AWMDM of the $\tau$ lepton. \label{values}}
\begin{tabular}{cc}
\hline
\hline
Parameter&Value\\
\hline
\hline
$f$&2 --6 TeV\\
$t_\beta$&1-9 \\
$z_3$&$\sim 1$\\
$\delta_{23}$&$\leq 0.1$\\
$m_\eta$& 20 GeV\\
$\delta_{\nu}$&$\leq 0.09$\\
$\sin2\theta$&$\leq 0.1$\\
\hline
\hline
\end{tabular}
\end{center}
\end{table}

In order to estimate the $\tau$ AMDM we used the Ma\-the\-matica numerical routines  to evaluate the parametric integrals involved in our calculation. A cross-check was done by evaluating the results expressed in terms of Passarino-Veltman scalar functions  via the numerical  FF/LoopTools routines \cite{vanOldenborgh:1989wn,Hahn:1998yk}. As already mentioned, it is convenient to analyze the behavior of the AMDM and AWMDM as  functions of the symmetry breaking scale $f$ since the mass of the new particles and the corrections to the SM couplings and particle masses  depend on it. Also, since the mixing angle $\delta_\nu$ depends on $t_\beta$, it is  worth examining the dependence on this parameter in the allowed interval.

\subsection{Anomalous magnetic dipole moment of the $\tau$ lepton}

\begin{figure*}[!hbt]
\centering
\subfigure[]{\includegraphics[width=8.25cm]{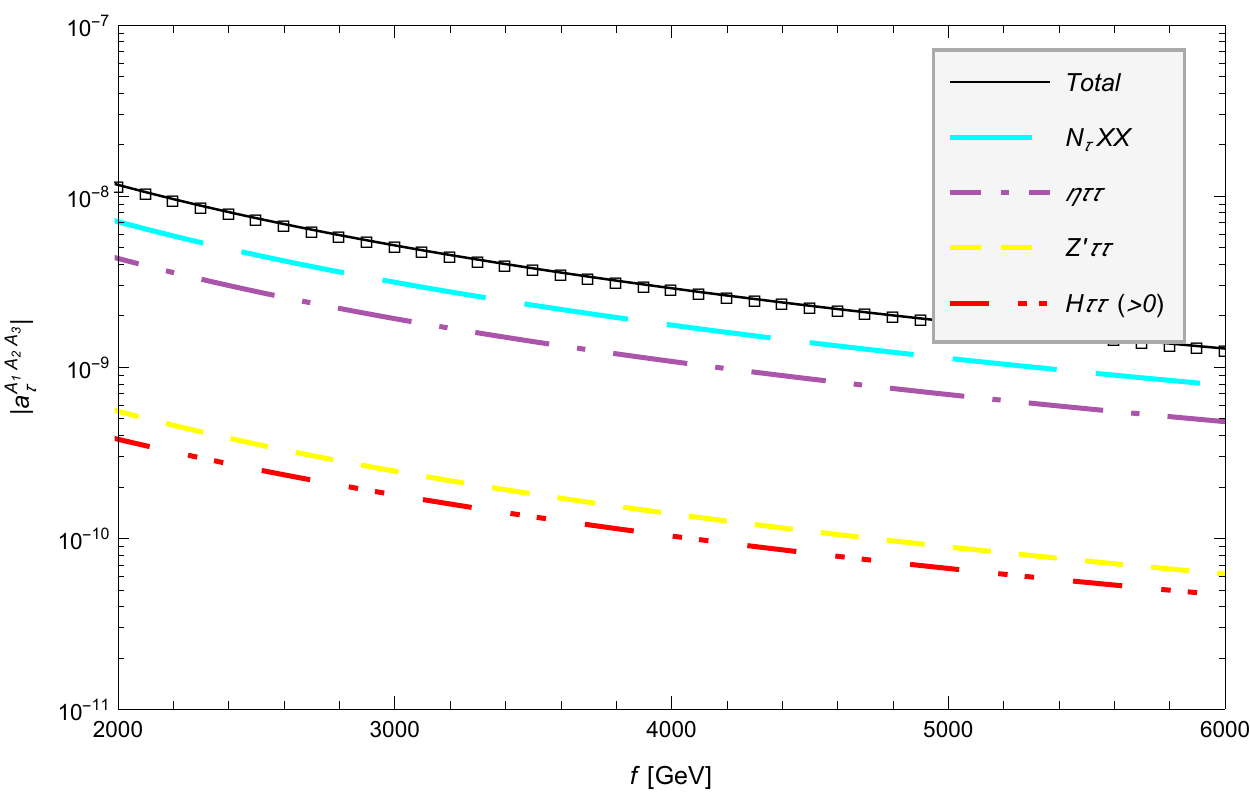}}
\subfigure[]{\includegraphics[width=8.25cm]{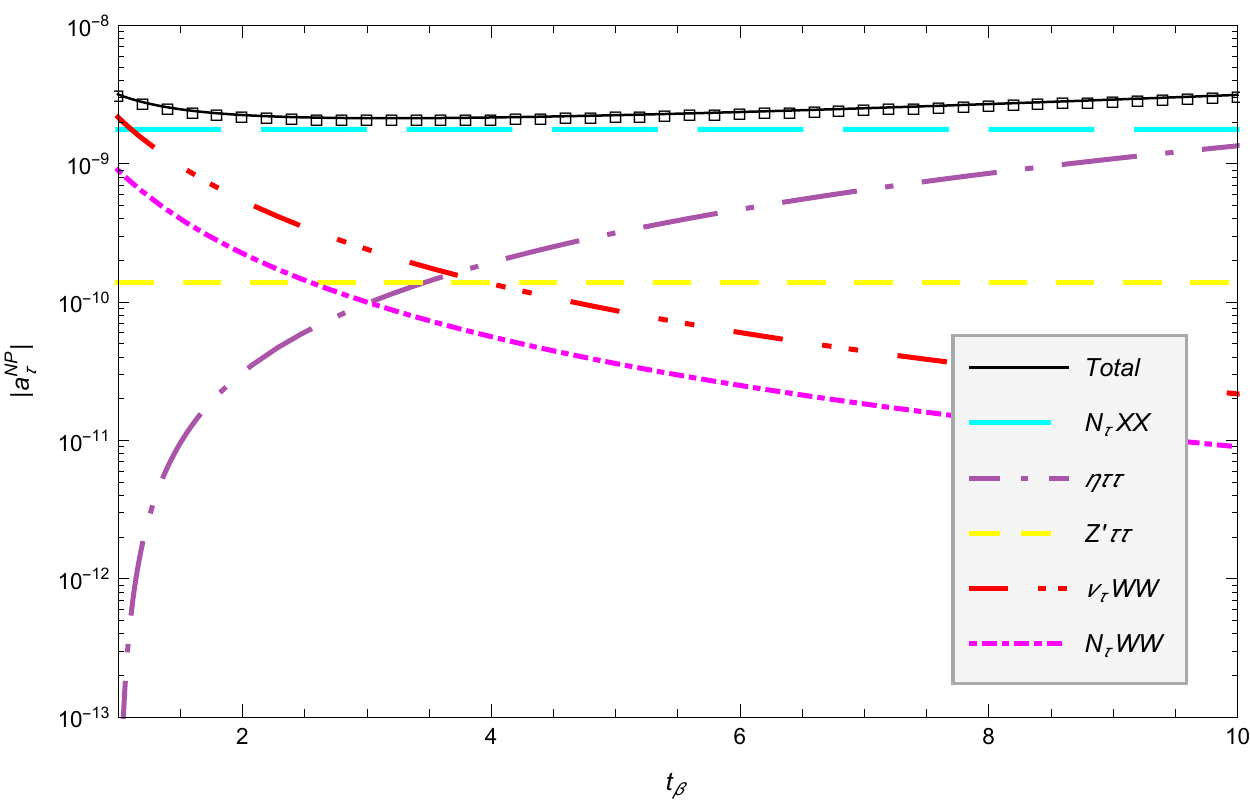}}
\caption{Absolute values of the main partial  contributions from the SLHM to $a_\tau^{NP}$ as functions of $f$ for $t_\beta=9$ (left plot) and as functions of $t_\beta$ for $f=4$ TeV (right plot). For the remaining parameters of the model we use the values shown in Table \ref{values}. The partial contributions below the $10^{-10}$ level are not shown. The three-letter tags denote the virtual particles circulating in each type of Feynman diagram. All the contributions are negative except the $H\tau\tau$ one. The absolute value of the sum all of the contributions is also shown (solid lines with squares).
\label{plotMDM}}
\end{figure*}

In the left plot of Fig. \ref{plotMDM} we show  the absolute values of  the main partial contributions to $a_{\tau}^{NP}$  along with the total sum as a function of $f$ for  $t_\beta=9$, whereas in the right plot we set $f=4$ TeV and show the dependence of $a_{\tau}^{NP}$ on $t_\beta$. For the remaining parameters  we use the values shown in Table \ref{values}.  We have refrained from showing the curves for the most suppressed contributions.

We first discuss the behavior observed in Fig. \ref{plotMDM}(a). Notice that the magnitude of each contribution depends highly on the respective $f_\gamma^{A_1 A_2 A_3}$ coefficient and to a lesser extent on the magnitude of the loop integral, which in turn dictates its behavior. Therefore, the $\nu_\tau XX$, $N_\tau WW$, and $\nu_\tau WW$ contributions, which are not shown in the plot, are the most suppressed ones, with values below the $10^{-10}$ level. This stems from the fact that the $f_\gamma^{A_1 A_2 A_3}$ coefficients associated with these contributions include two powers of the coupling constants $g_L^{Vn\ell}$, which  are of the order of $\delta_\nu\sim v/f$, thereby being considerably suppressed for large $f$. Although the $H\tau\tau$  and $Z'\tau\tau$ contributions are less suppressed, they  are below the $10^{-9}$ level, whereas the $\eta\tau\tau$ and  $N_\tau XX$ contributions are the largest ones and  can reach values up to the order of $10^{-8}$ for $f$ around 2 TeV,  which is a result of the fact that  the respective $f_\gamma^{A_1 A_2 A_3}$ coefficients have no $(v/f)^2$ suppression factor. Another point worth to mention is that all the partial contributions  are negative except for the $H\tau\tau$ and $N_\tau WW$ ones. Since these contributions are relatively small, they will not interfere with the dominant contributions, which will add up constructively. In conclusion both the $\eta\tau\tau$ and  $N_\tau XX$ contributions will represent the bulk of the total contribution to $a_{\tau}^{NP}$,  which is  of the order of $10^{-8}$ for $f=2$ TeV, but has a decrease of about one order of magnitude  as $f$ increases  up to 6 TeV, as observed in the plot.

We now turn to discuss the dependence of $a_{\tau}^{NP}$ on the $t_\beta$ parameter as depicted in Fig. \ref{plotMDM}(b). We  observe that the $N_\tau XX$ contribution, which has a very slight dependence on $t_\beta$ indeed, is the dominant one, with marginal contributions arising from other diagrams. In  the allowed $t_\beta$ interval, the $H\tau\tau$ contribution is negligible and is  not shown in the plot. For low $t_\beta$, the $\nu_\tau WW$ and $N_\tau WW$ contributions can be as large as the $N_\tau XX$ one, but the $\eta\tau\tau$ contribution  is the one that becomes important when $t_\beta$ increases. Since these contributions are directly proportional to the square of the mixing parameter $\delta_\nu=v/(\sqrt{2}t_\beta f)$, they get suppressed by two orders of magnitude as $t_\beta$ goes from 1 to 9. We observe that the total contribution of the SLHM to $a_{\tau}^{NP}$ remains almost constant in this $t_\beta$ interval as it is dominated by the $N_\tau XX$ contribution.

\begin{figure}[!hbt]
\centering{\includegraphics[width=8.25cm]{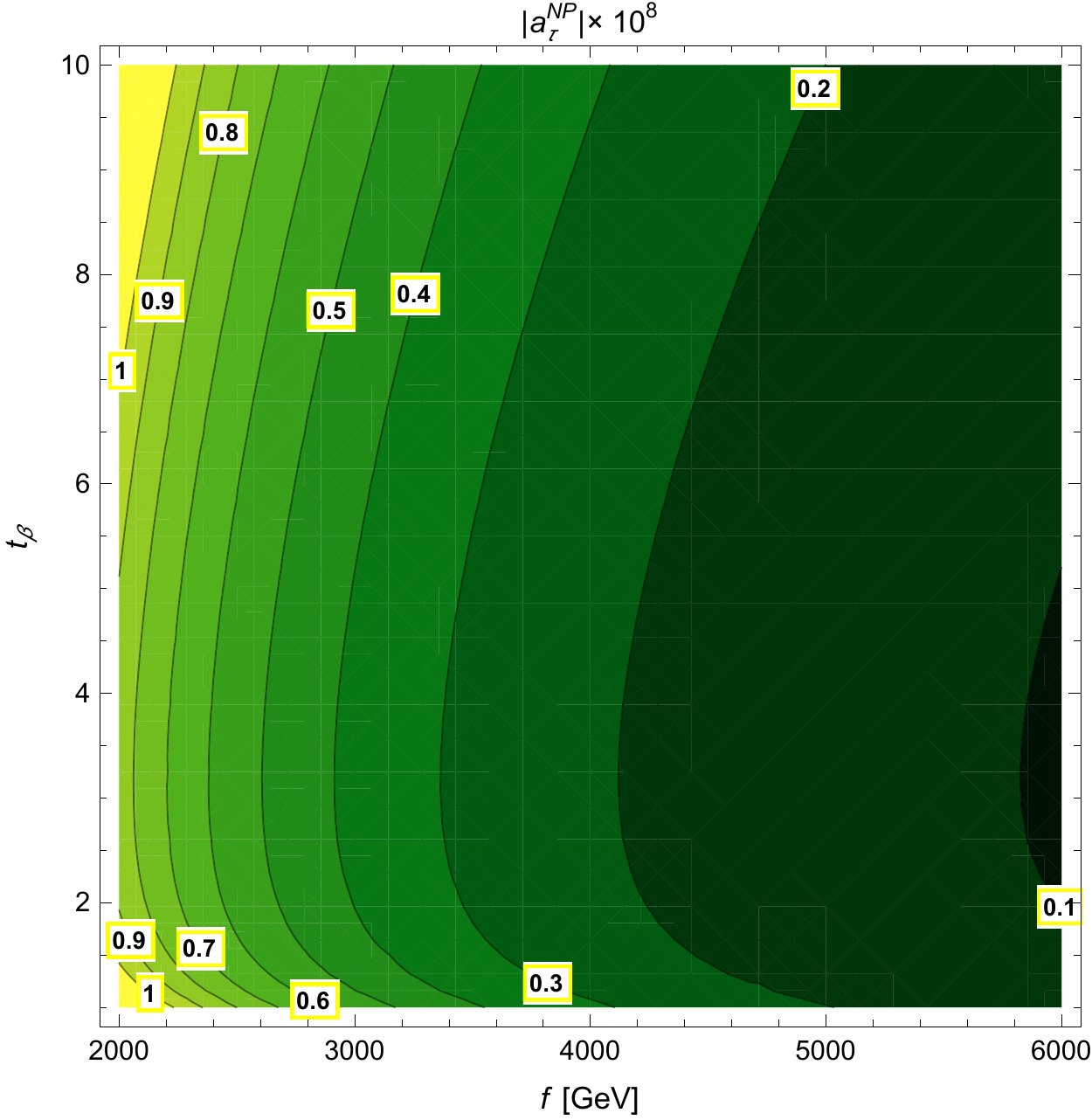}}
\caption{Contours of the SLHM contribution to $|a_\tau^{NP}|$ in the $f$ vs $t_\beta$ plane. For the remaining parameters of the model we use the values shown in Table \ref{values}. \label{contplotMDMTot}}
\end{figure}

The behavior of the SLHM contribution to $a_\tau^{NP}$ is best illustrated in Fig. \ref{contplotMDMTot}, where we plot the contours of $|a_\tau^{NP}|$ in the $f$ vs $t_\beta$ plane for the parameter values of Table \ref{values}.  We observe that the SLHM contribution to $a_\tau^{NP}$ is of the order of  $10^{-8}$ for $f$ between 2 TeV and 5 TeV, but it is below $10^{-9}$ for $f$ above $5$ TeV and decreases rapidly as $f$ increases.
So, if we consider the most stringent constraint on $f$, namely 5.6 TeV, we can expect values of $a_\tau^{NP}$ of the order of $10^{-9}$. We also observe that there is little dependence of  $a_\tau^{NP}$ on the value of $t_\beta$, but such dependence is more pronounced for large $f$. It is interesting to make a comparison with the  typical predictions of some popular extension models as reported in the literature.
In this respect, several extension models predict values for $a_{\tau}^{NP}$ lying  in the interval between $10^{-9}$ and $10^{-6}$ \cite{Bolanos:2013tda,Moyotl:2012zz,Ibrahim:2008gg,Arroyo-Urena:2015uoa}.   We note that although the SLHM  contribution  is of the same order than the potential contribution of leptoquark models (LQM) \cite{Bolanos:2013tda}, it  is disfavored with respect to the contributions of THDMs \cite{Bernabeu:1995gs,GomezDumm:1999tz}, the minimal supersymmetric standard model (MSSM) \cite{Hollik:1998wk,Hollik:1998vz}, and unparticles  (UP)\cite{Moyotl:2012zz}, which can reach values as high as $10^{-6}$.

\subsection{Anomalous weak magnetic dipole moment of the $\tau$ lepton}

\begin{figure*}[!hbt]
\centering
\subfigure[]{\includegraphics[width=8.25cm]{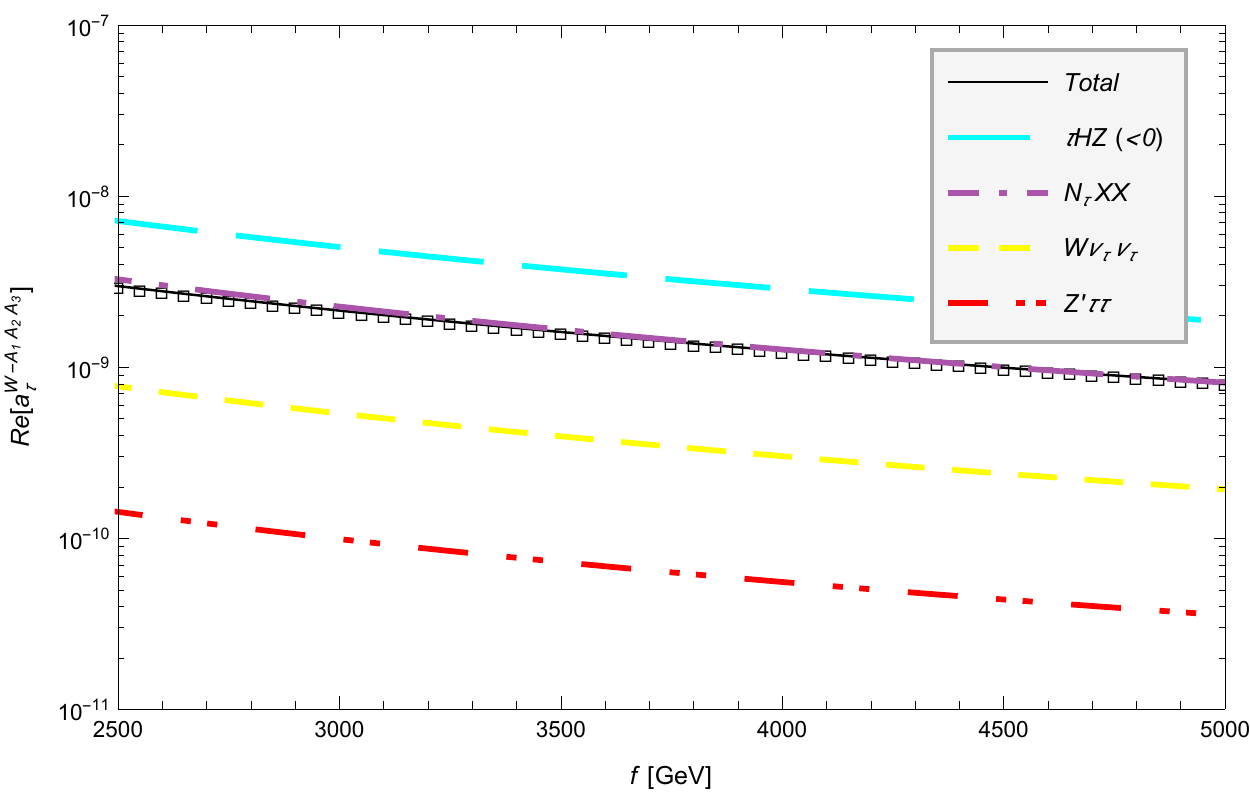}}
\subfigure[]{\includegraphics[width=8.25cm]{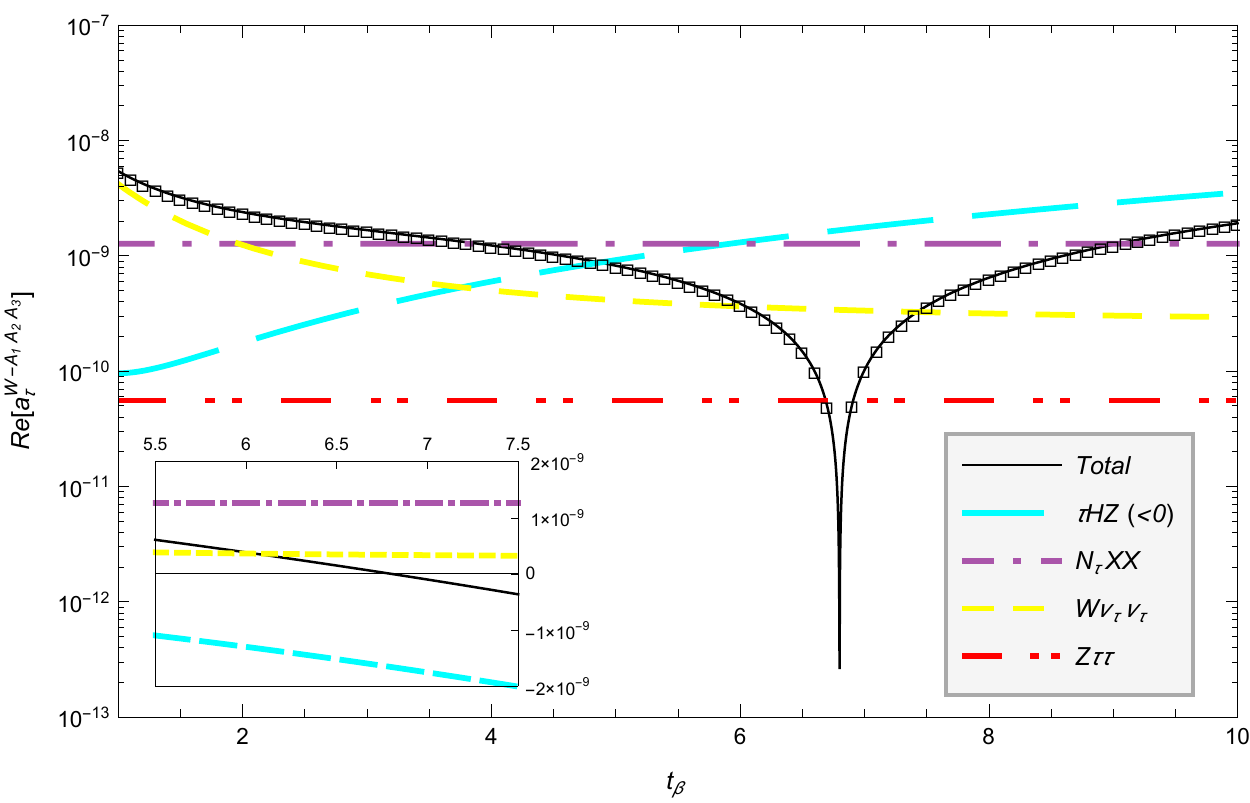}}
\caption{Absolute value of the real part of the main partial contributions to $a_{\tau}^{W-NP}$ and the total sum as  a function of $f$  for $t_\beta=9$ (left plot) and as a function of $t_\beta$ for $f=4$ TeV. For the remaining parameters of the model we use the  values of Table \ref{values}. The contributions below the $10^{-10}$ level are not shown. All the contributions are positive except the $\tau h Z$ one. The absolute value of the sum all of the contributions is also shown (solid lines with squares).  In the  bottom-right corner we zoom into the region  where $a_{\tau}^{W-NP}$ changes from positive to negative due to the cancellation between the distinct partial contributions. \label{plotReWMDM} }
\end{figure*}

We  now present the analysis of the NP contribution of the SLHM to the AWMDM of the $\tau$ lepton. There are some differences  with respect to the behavior of the AMDM: apart that  the AWMDM receives extra contributions, it can develop an imaginary part, which  arises from the diagrams where the external $Z$ gauge boson is attached to a couple of particles whose total mass is lower than $m_Z$. This occurs when there are  two internal SM neutrinos or  $\tau$ leptons in the loop.  Again, the magnitude of each partial contribution to the   AWMDM is highly dependent on the corresponding $f_Z^{A_1A_2A_3}$ coefficient, while its behavior is dictated by the loop integral. We  show in Fig. \ref{plotReWMDM} the real part of the dominant partial contributions of the SLHM to  $a_\tau^{W-NP}$ as well as the total sum as functions of $f$ for $t_\beta=9$ (left plot) and as functions of $t_\beta$ for $f=4$ TeV (right plot). For the remaining parameters we use the same set of values used in our analysis of the  AMDM. All the contributions not shown  in the plots are negligible.

We first discuss the behavior of ${\rm Re}[a_\tau^{W-NP}]$ as a function of $f$ as shown Fig. \ref{plotReWMDM}(a). Numerical evaluation shows that for $f$ around  2 TeV the magnitude of the partial contributions   ranges from $10^{-14}$ to $10^{-9}$, with an additional suppression of at least one order of magnitude  for $f$ around 5 TeV. The most suppressed contributions are
the $\nu_\tau XX$, $X\nu_\tau\nu_\tau$ and $W N_\tau N_\tau$ ones, which are  proportional to the $(v/f)^2$ factor and  are    below the $10^{-13}$ level. Other contributions such as $\tau HZ'$, $\nu_\tau WW$, $WN_\tau \nu_\tau$, $N_\tau WW$, $\eta \tau\tau$, $H \tau\tau$, and  $XXN_\tau$ are less suppressed but are also below the $10^{-10}$ level.
In fact, only the contributions shown in the plot,  $\tau HZ$, $N_\tau XX$, and $W\nu_\tau \nu_\tau$, are relevant for the total sum. While the $\tau HZ$  contribution  is the dominant one,  the $N_\tau XX$ contribution plays a subdominant role, which again   is due to the fact that the $f^{N_\tau XX}_Z$ coefficient is not suppressed by the $(v/f)^2$ factor. We also observe that the  $W\nu_\tau\nu_\tau$ contribution, which together with the $\tau HZ$ contribution  are absent in the AMDM, is below the $10^{-9}$ level. Very interestingly, the $H\tau\tau$ and $\eta\tau\tau$ contributions, which were not very suppressed  in the AMDM case, are now negligible as  they are proportional to the small $g_{V}^{Z\ell\ell}$ coupling.  Note also that all the contributions shown in the plot are positive except for the $\tau HZ$ one. As a result   the total contribution will have an additional suppression as the main contributions  will have a large cancelation due to their opposite signs. This becomes evident in the curve for the total contribution, which appears below the curve for the $\tau HZ$ contribution.

As far as the behavior of the real part of  $a_\tau^{W-NP}$ as a function of $t_\beta$ is concerned , we observe in Fig. \ref{plotReWMDM}(b) that the $\tau HZ$ and $W\nu_\tau\nu_\tau$ contributions are highly dependent on $t_\beta$: in the interval  where this parameter increases from 1 to 10, the $\tau HZ$ contribution is negative and  increases by one order of magnitude, whereas the $W\nu_\tau\nu_\tau$ contribution is positive and decreases  by one order of magnitude. On the other hand, the $N_\tau XX$ contribution is positive and remains almost constant throughout this $t_\beta$ interval. It is interesting  that due to the opposite signs of the partial contributions, there is a flip of sign of the total contribution around $t_\beta=6.8$.  For low $t_\beta$,   the total contribution is positive as it  arises mainly from the $N_\tau XX$ and $W\nu_\tau\nu_\tau$  contributions, with the $\tau H Z$ contribution being subdominant. As $t_\beta$ increases,  the magnitude of the $\tau H Z$  contribution increases, whereas that of the $W\nu_\tau\nu_\tau$  contribution decreases. Therefore,  the total sum cancels out around $t_\beta=6.8$ and   becomes negative above this value since the $\tau HZ$ contribution becomes the dominant one. This effect is evident in the large dip of  the total contribution curve, which is due to the flip of sign of the AWMDM. This behavior of the partial and total contributions around $t_\beta=6.8$ is best illustrated in the zoomed region displayed  at the bottom-right corner of Fig. \ref{plotReWMDM}(b), where we show the  $\tau$ AWMDM contributions without taking their absolute values.

\begin{figure}[!hbt]
\centering{\includegraphics[width=8.25cm]{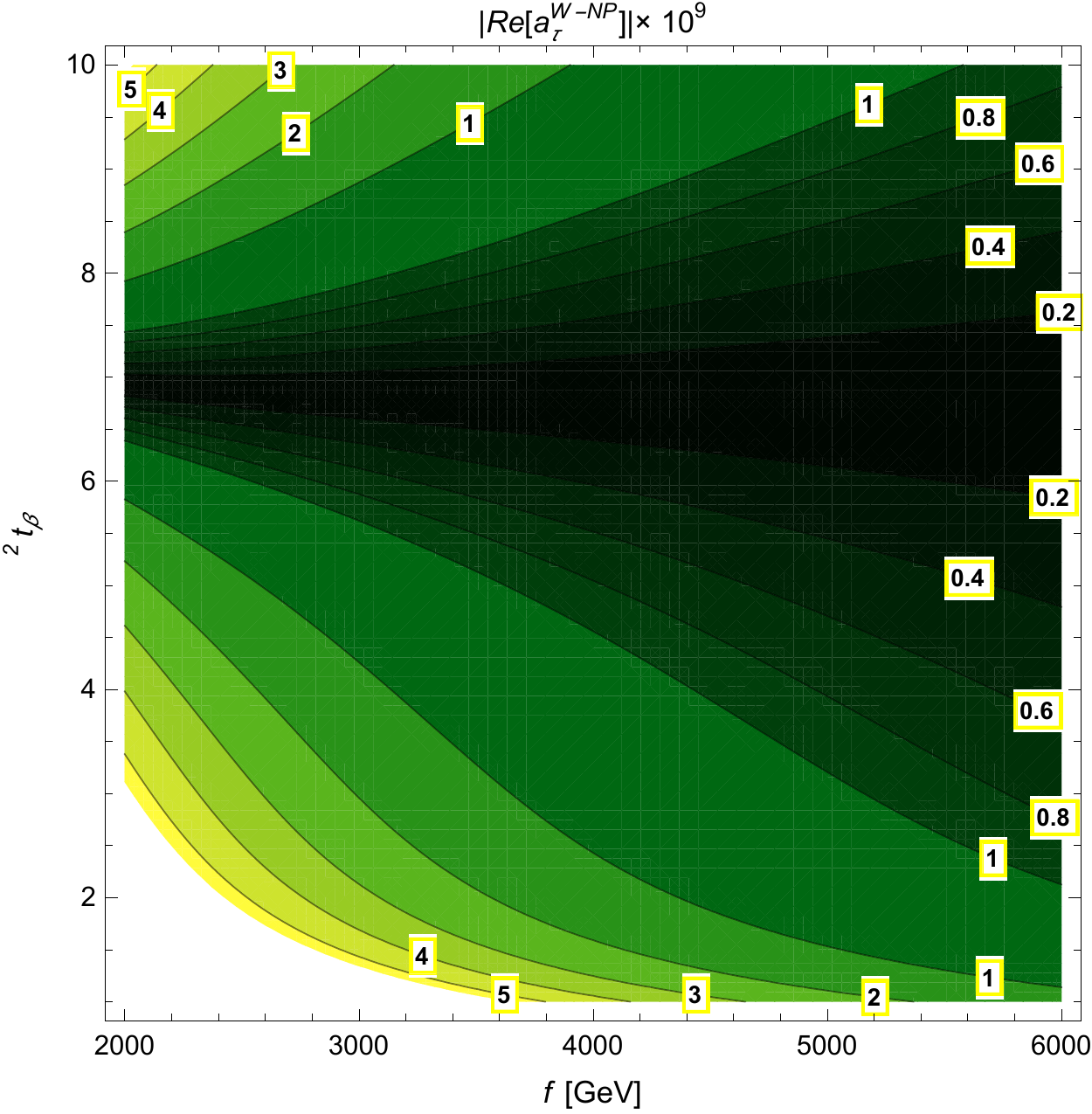}}
\caption{Contours of the SLHM contribution to the absolute value of the real part of $a_\tau^{W-NP}$ in the $f$ vs $t_\beta$ plane. For the remaining  parameters of the model we use the values shown in Table \ref{values}. \label{contplotReWMDMTot}}
\end{figure}

The behavior of the real part of $a_\tau^{W-NP}$ as a function of $f$ and $t_\beta$ is best illustrated  in Fig. \ref{contplotReWMDMTot}, where we show the contours of the real part of the total SLHM contribution to $a_\tau^{W-NP}$ in the $f$ vs $t_\beta$ plane. It can be observed that the real part of $a_\tau^{W-NP}$ reaches its largest values, of the order of $10^{-9}$, for low and high $t_\beta$, irrespective of the value of $f$. There is a band centered around $t_\beta\sim 7$ where the lowest values of real part of $a_\tau^{W-NP}$ are reached. Such a band (darkest region) widens  as $f$ increases . We observe that for $t_\beta$ around  10, there is a slow decrease of ${\rm Re}[a_W^{W-NP}]$ as $f$ increases but in general its magnitude is of the order of $10^{-10}-10^{-9}$. A comparison of the SLHM contribution with  typical predictions of some SM extensions allow us to conclude that for $f$ up to $4$ TeV, the SLHM contribution can be above the $10^{-9}$ level and  can  be  larger than the contributions predicted by THDMs, the MSSM and UP, which are of the order of $10^{-10}$. For $f\gtrsim 4$ TeV, the SLHM contribution decreases and it is expected to be smaller than the values predicted by other extension models.

\begin{figure*}[!hbt]
\centering
\subfigure[]{\includegraphics[width=8.25cm]{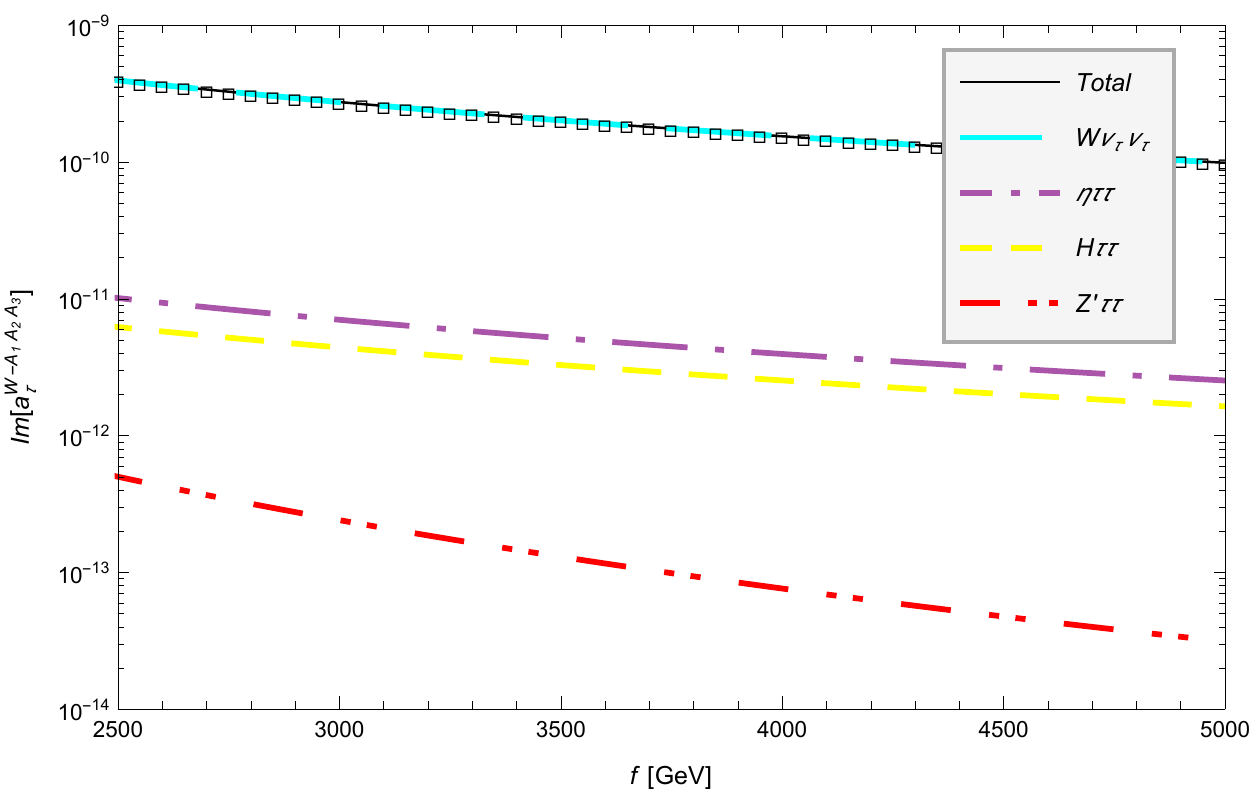}}
\subfigure[]{\includegraphics[width=8.25cm]{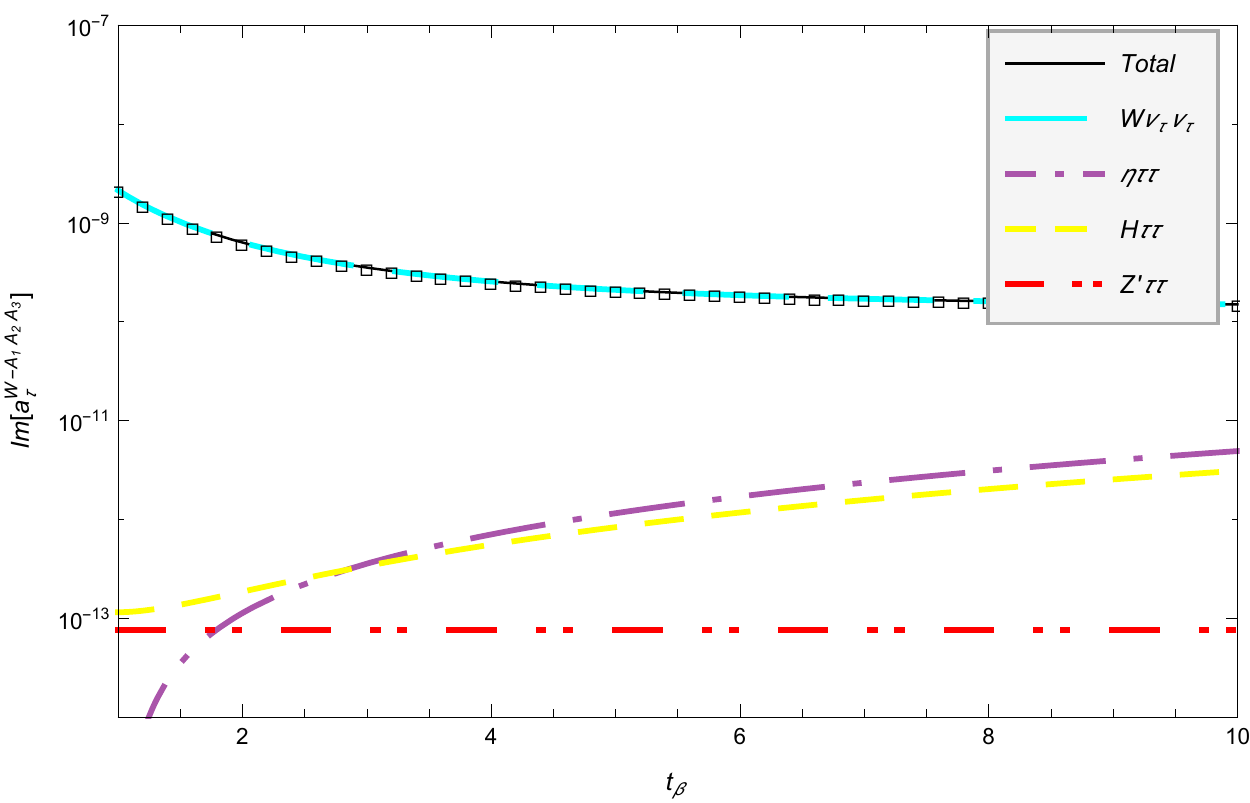}}
\caption{The same as in Fig. \ref{plotReWMDM} but for the imaginary part  of $a_{\tau}^{W-NP}$.   \label{plotImWMDM} }
\end{figure*}

We now turn to examine the behavior of the imaginary part of the partial contributions of the SLHM to $a_\tau^{W-NP}$. We will follow the same approach as that used for the analysis of the real part. There are only four  contributions to $a_\tau^{W-NP}$ that can develop an imaginary part. In Fig. \ref{plotImWMDM}(a) we show the behavior of such  contributions as a function of $f$ for $t_\beta=9$. Contrary to what happens with the real part, the imaginary part of the $W\nu_\tau \nu_\tau$ contribution is the dominant one by far, at the level of $10^{-10}$, with the imaginary parts of the $\eta \tau\tau$,  $H\tau\tau$, and $Z'\tau\tau$ contributions suppressed by more than one order of magnitude. Therefore, the imaginary part of  $a_\tau^{W-NP}$ will be completely dominated by the $W\nu_\tau \nu_\tau$ contribution. In fact both the curve for the $W\nu_\tau \nu_\tau$ contribution and  the total contribution overlap. In this region of the parameter space of the SLHM, the imaginary part of $a_\tau^{W-NP}$ is positive, with a magnitude of the order of $10^{-10}$, which slightly decreases as $f$ increases. As far as the behavior of the imaginary part of $a_{\tau}^{W-NP}$ as a function of $t_\beta$ is concerned [Fig. \ref{plotImWMDM}(b)], there is no considerable change in the analysis as in the allowed $t_\beta$ interval  the imaginary part of the $W\nu_\tau  \nu_\tau$ contribution is dominant, whereas the remaining contributions are negligibly small. For low $t_\beta$, the imaginary part of the total contribution to $a_{\tau}^{W-NP}$ can be of the order of $10^{-9}$, but it decreases by almost one order of magnitude as $t_\beta$ goes up to 10.

\begin{figure}[!hbt]
\centering{\includegraphics[width=8.25cm]{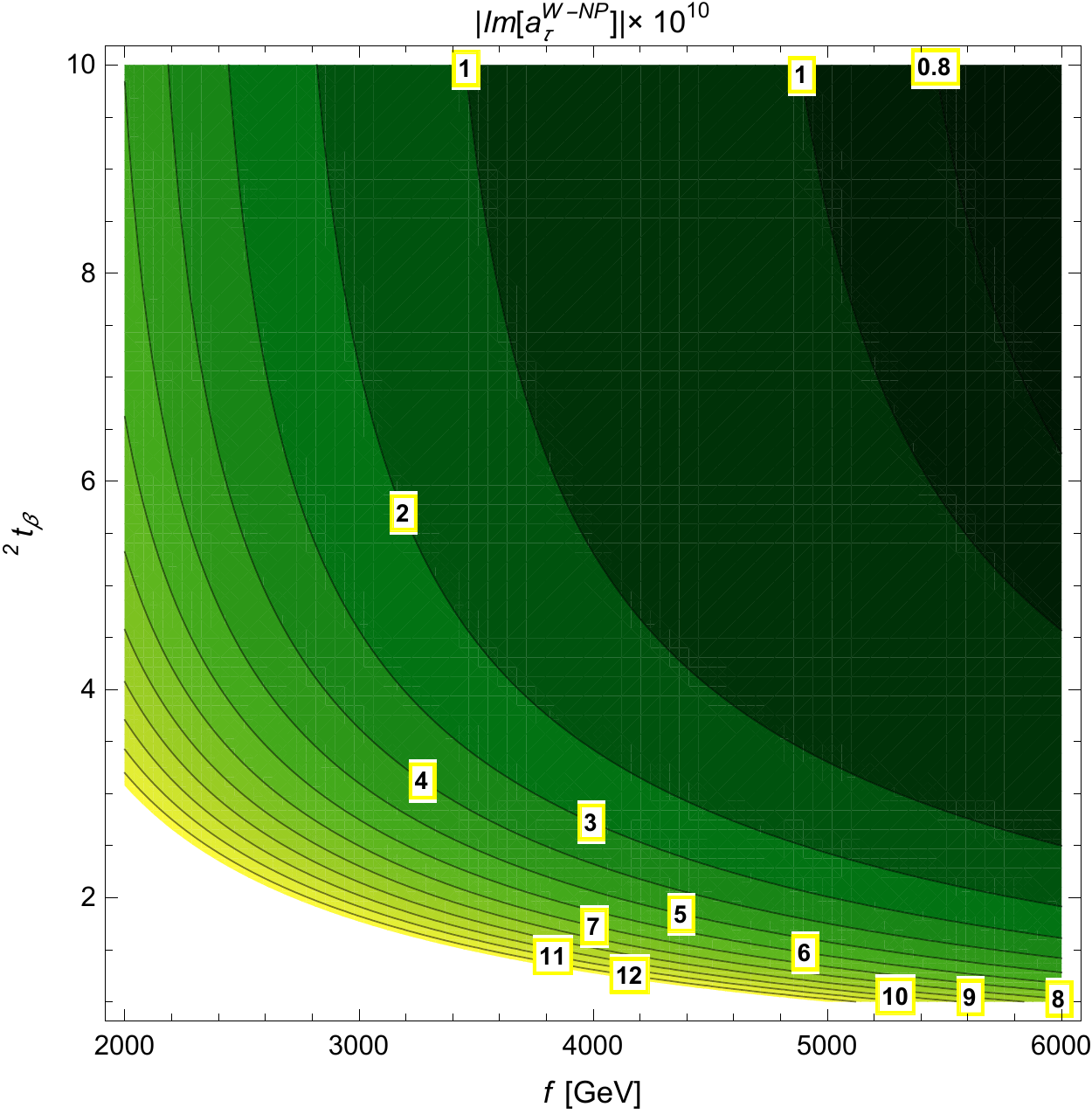}}
\caption{The same as in Fig. \ref{contplotReWMDMTot} but for the imaginary part  of $a_{\tau}^{W-NP}$.    \label{contplotImWMDMTot}}
\end{figure}

Finally we present the contours of the imaginary part  of $a_{\tau}^{W-NP}$ in the $f$ vs $t_\beta$ plane in Fig. \ref{contplotImWMDMTot}. We observe that the imaginary part  of $a_{\tau}^{W-NP}$ can be of the order of $10^{-9}$ for low values of $t_\beta$, irrespective of the value of $f$. This is also true  for $t_\beta\sim 10$  and $f\le 3$ TeV, but there is a pronounced decrease of about one order of magnitude for larger values of  $f$. We also can observe that  ${\rm Im}[a_\tau^{W-NP}]$ decreases mildly as $f$ increases.  As far as  the values predicted by other extension models, although the imaginary part of the total contribution of the SLHM  is larger than that predicted by  type-I and type-II LQMs (of the order of $10^{-10}$) it is well below the contributions predicted by THDMs and the MSSM (of the order of $10^{-7}$).

\section{Conclusions}\label{conclusions}
In this work we have calculated analytical expressions, both in terms of parametric integrals and Passarino-Veltman scalar functions, for the one-loop contributions to the static anomalous magnetic and weak magnetic  dipole moments of a char\-ged lepton in the context of the SLHM. We have considered the scenario in which there is no $CP$ violation in the model and thereby  there are no electric nor weak electric dipole moments. The expressions presented for the weak properties are very general and can be useful to compute the weak properties of a charged lepton in other extension models.  For the numerical analysis we have focused on the case of the $\tau$ lepton since their electromagnetic and weak properties are the least studied in the literature and also because they have great potential  to  be experimentally tested in the future. For values of the parameters of the model allowed by current experimental data we find that the respective contribution to the  $\tau$ AMDM  is of the order of $ 10^{-9}$, whereas the real (imaginary) part of the $\tau$ AWMDM  is of the order of $10^{-9}$ ($10^{-10}$). The  SLHM contribution to the $\tau$ AMDM could have some enhancement  in the scenario in which there is a very light pseudoscalar boson $\eta$, with a mass of the order of about $10$ GeV, and  $t_{\beta}$ is of the order of 20. However  such a  value of $t_\beta$ are already excluded according to the bounds obtained from experimental data. Proposed future experiments are expected to reach a sensitivity to the $\tau$ AMDM and the real part of the AWMDM of the order of $10^{-6}$ \cite{Bernabeu:2007rr,Fael:2013ij,Eidelman:2016aih}  and $10^{-4}$ \cite{Bernabeu:1994wh}. Therefore, the values predicted for these observables by the SLHM would be out of the reach of experimental detection.

One further remark is in order here. Little Higgs models are effective
theories valid up to the cut-off scale $\Lambda= 4 \pi f$. Below this
scale a good prediction is provided by the effective theory, but at
higher energies the physics would become strongly coupled and the
effective theory must be replaced by its ultraviolet (UV)
completion, which would be a QCD-like gauge theory with a
confinement scale around 10 TeV (see for instance \cite{Katz:2003sn}). This gives rise to the possibility
that the EWSB is driven by strong dynamics such as occurs in
technicolor theories. It is thus possible that the $\tau$ AMDM and
AWMDM can receive some enhancement from the UV completion, but an analysis along these lines is beyond the scope of the present work.


\begin{acknowledgements}
We acknowledge financial support from Sistema Nacional de Investigadores (Mexico), Consejo Nacional  de Ciencia y Tecnolog\'ia (Mexico) and Vicerrector\'ia de  Investigaci\'on y Estudios de Posgrado (BUAP).
\end{acknowledgements}

\onecolumn
\appendix

\section{Feynman Rules in the SLHM}
\label{FeynmanRules}

We now present the Feynman rules necessary for our calculation (see \cite{delAguila:2011wk} for a complete set of SLHM Feynman rules). We note that the coupling constant associated with the  $A_1A_2A_3$ vertex will be written as $ieg^{A_1A_2A_3}$, so the $f_V^{A_1A_2A_3}$ ($V=Z,\gamma$) coefficients of Eqs. (\ref{a^W_ell}) and (\ref{a_ell}) will be given in terms of the $g^{A_1A_2A_3}$ constants.   In Table \ref{triplevertices} we show the Feynman rules for the trilinear gauge boson couplings $V_i V_j^+V_j^-$, whereas in Table \ref{Vff} we show the ones  for the vertices of a gauge boson to a fermion pair $V\bar{f_i}f_j$. Finally, Table \ref{scalarvertices} gathers all the Feynman rules for the scalar interactions.

\begin{table}[htb!]
\renewcommand{\arraystretch}{1.6}
\begin{center}
\caption{Feynman rules for the trilinear gauge boson couplings $V_{\mu}(p_{1})V^+_{\nu}(p_{2})V^-_{\rho}(p_{3})$. The Feynman rule for all these couplings have the generic form $ieg^{V_i V_j V_j}\left(g_{\mu\nu}(p_{2}-p_{1})_{\rho}+g_{\nu\rho}(p_{3}-p_{2})_{\mu}+g_{\mu\rho}(p_{1}-p_{3})_{\nu}\right)$, where the $g^{V_i V_j V_j}$ constants are shown in the right column. All the four-momenta are taken incoming. \label{triplevertices}}
\begin{tabular}{cc}
\hline
\hline
{{$V_i V_j V_j$ vertex}} & {{$g^{V_i V_j V_j}$}}\\
\hline
\hline
{{$A X^{+}X^{-}$}} &  $-1$ \\
{{$A W^{+}W^{-}$}} &  $-1$ \\
$ZX^{+}X^{-}$ & $\frac{c_{W}^{2}-s_{W}^{2}}{2s_Wc_{W}}-\frac{\delta_{Z}}{2}\frac{\sqrt{3-t_{W}^{2}}}{s_W}$\\
$ZW^{+}W^{-}$ & $\frac{c_{W}}{s_W}$\\
\hline
\hline
\end{tabular}
\end{center}
\end{table}

\begin{table*}[htb!]
\renewcommand{\arraystretch}{1.6}
\begin{center}
\caption{Feynman rules for the vertices of a gauge boson to a lepton pair $V_\mu\bar{f_i}f_j$. The Feynman rule for this class of vertices has the generic form $ie(g_{L}^{V\bar{f_i} f_j}P_{L}+g_{R}^{V\bar{f_i} f_j}P_{R})\gamma_\mu$, with $P_L$ and $P_R$ the left- and right-chirality projectors, where the $g_{L}^{V\bar{f_i} f_j}$ and $g_{R}^{V\bar{f_i} f_j}$  constants are shown in the second and third columns.  The vector  and axial couplings are
$2g_{V}^{V\bar{f_i} f_j}=g_{L}^{V\bar{f_i} f_j}+g_{R}^{V\bar{f_i} f_j}$ and $2g_{A}^{V\bar{f_i} f_j}=g_{L}^{V\bar{f_i} f_j}-g_{R}^{V\bar{f_i} f_j}$.
\label{Vff}}
\begin{tabular}{ccc}
\hline
\hline
$V_\mu\bar{f_i}f_j$ vertex& $g_{L}^{V\bar{f_i} f_j}$ &$g_{R}^{V\bar{f_i} f_j}$\\
\hline
\hline
$Z\bar{\ell}\ell$ & $\frac{-1+2s_{W}^{2}}{2s_Wc_{W}}+\frac{1-2s_{W}^{2}}{2s_Wc_{W}^{2}\sqrt{3-t_{W}^{2}}}\delta_{Z}$ & $\frac{s_{W}}{c_{W}}-\frac{s_{W}}{c_{W}^{2}\sqrt{3-t_{W}^{2}}}\delta_{Z}$\\
$Z\bar{\nu}\nu$ & $\frac{1}{2s_Wc_{W}}(1-\delta_{\nu}^{2})+\frac{1-2s_{W}^{2}}{2s_Wc_{W}^{2}\sqrt{3-t_{W}^{2}}}\delta_{Z}$ & $0$\\
$Z\bar{N}N$ & $\frac{1}{2s_Wc_{W}}\delta_{\nu}^{2}-\frac{1}{s_W\sqrt{3-t_{W}^{2}}}\delta_{Z}$ & $0$\\
$Z\bar{N}_{m}\nu$ & $-\frac{1}{2s_Wc_{W}}\delta_{\nu}V_{l}^{mi}$ & $0$\\
$W^{+}\bar{\nu}\ell$ & $\frac{1}{\sqrt{2}s_W}\left(1-\frac{\delta_{\nu}^{2}}{2}\right)$ & $0$\\
$W^{+}\bar{N}_{m}\ell$ & $-\frac{1}{\sqrt{2}s_W}\delta_{\nu}V_{l}^{mi}$ & $0$\\
$X^{+}\bar{\nu}\ell$ & $-\frac{i}{\sqrt{2}s_W}\delta_{\nu}$ & $0$\\
$X^{+}\bar{N}_{m}\ell$ & $-\frac{i}{\sqrt{2}s_W}\left(1-\frac{\delta_{\nu}^{2}}{2}\right)V_{l}^{mi}$ & $0$\\
$Z'\bar{\ell}\ell$ & $\frac{1-2s_{W}^{2}}{2s_Wc_{W}^2\sqrt{3-t_{W}^{2}}}+\frac{1-2s_{W}^{2}}{2s_Wc_{W}}\delta_{Z}$ & $-\frac{s_W}{c_W^2\sqrt{3-t_{W}^{2}}}-\frac{s_W}{c_W}\delta_{Z} $\\
\hline
\hline
\end{tabular}
\end{center}
\end{table*}

\begin{table*}[htb!]
\renewcommand{\arraystretch}{1.6}
\begin{center}
\caption{Feynman rules for the vertices involving the scalar bosons $H$ and $\eta$ at the leading order in $v/f$ in the  SLHM.  \label{scalarvertices}}
\begin{tabular}{ccc}
\hline
\hline
$A_1 A_2 A_3$ vertex&Feynman rule&$g^{A_1A_2A_3}$\\
\hline
\hline
$H\bar{\ell}\ell$ &$ie g^{H\bar{\ell}\ell}$&$-\frac{m_{\ell}}{2s_Wm_W}\left(1-\frac{v^2}{6f^2}\left(\frac{1}{t_\beta}+t_\beta\right)^2\right)$ \\
$H  Z_\alpha Z_\beta $ &$ieg^{ZZH}g_{\alpha\beta}$&$\frac{gv}{2c_W s_W}\left(1-\frac{v^2}{4f^2}\left(\frac{1}{3}
\left(\frac{s_\beta^4}{c_\beta^2}+\frac{c_\beta^4}{s_\beta^2}\right)
+\frac{1}{4}\left(1-t_W^2\right)^2\right)\right)$\\
$\eta\bar{\ell}\ell$ &$ieg^{\eta\bar{\ell}\ell}\gamma^5 $&$\frac{im_{\ell}}{\sqrt{2}s_Wf} \left(\frac{1}{t_\beta}-t_\beta\right)$\\
$H  Z_\alpha Z'_\beta $&$ieg^{ H Z  Z' } g_{\alpha\beta}$&$\frac{gv(1-t_W^2)}{2s_Wc_W\sqrt{3-t_W^2}} $\\
\hline
\hline
\end{tabular}
\end{center}
\end{table*}

\section{Loop integrals}
\label{LoopIntegrals}

The AWMDM and AWMDM of a lepton are given by Eqs. (\ref{a^W_ell}) and (\ref{a_ell}). The $f^{A_1A_2A_3}_V$ coefficients  are shown in Table \ref{f^{A_1A_2A_3}} and the loop functions  $I_{V}^{A_1A_2A-3}$ ($V=Z,\gamma$)   will be presented below. The loop integration was performed via both    Feynman parametrization and the Passarino-Veltman method \cite{Passarino:1978jh}.  For the Dirac algebra and the Passarino-Veltman reduction  we used the Feyncalc routines \cite{Mertig:1990an}, and a further simplification was done with the help of the Mathematica symbolic algebra routines.  We will first present  the results  in terms of parametric integrals.

\begin{table}[hbt!]
\renewcommand{\arraystretch}{1.6}
\begin{center}
\caption{Coefficients $f^{A_1A_2A_3}_Z$   and $f^{A_1A_2A_3}_\gamma$ required in Eqs. (\ref{a^W_ell}) and (\ref{a_ell}), respectively, to compute the AWMDM and AMDM of a charged lepton arising from  loops carrying the $A_1A_2A_3$ particles. Here  $n$ is the SM neutrino $\nu$ or a heavy one $N$.  The coupling constants $g^{A_1A_2A_3}$ are shown in  \ref{FeynmanRules}. \label{f^{A_1A_2A_3}}}
\begin{tabular}{ccccc}
\hline
\hline
$A_1A_2A_3$&&$f^{A_1A_2A_3}_Z$ &&$f^{A_1A_2A_3}_\gamma$ \\
\hline
\hline
$nVV\quad (V=W,X)$&&$g^{Z VV} \left(g_{L}^{Vn\ell}\right)^{2}$&&$g^{A VV}\left(g_{L}^{V\ell n}\right)^{2}$\\
$Z'\ell\ell$&&$1$&&$1$\\
$Vnn\quad(V=W,X)$&&$ g_{L}^{Znn} \left(g_{L}^{Vn\ell}\right)^{2}$&&--\\
$V\nu N \quad(V=W,X)$&&$g_{L}^{ZN\nu}g_{L}^{V\nu\ell}g_{L}^{V N\ell}$&&--\\
$S\ell \ell\quad S=H,\eta)$&&$g_V^{Z\ell\ell}\left(g^{S\ell \ell}\right)^2$&&$\left(g^{S\ell\ell}\right)^2$\\
$\ell HV \quad (V=Z,Z')$&&$\frac{1}{m_Z} g^{HZV}g^{H\ell\ell} g_V^{V\ell\ell}$&&--\\
\hline
\hline
\end{tabular}
\end{center}
\end{table}

\subsection{Parametric integrals}

After introducing Feynman parameters and integrating over the 4-momentum space, the $I_V^{A_1A_2A_3}$ loop integrals can be cast in the following form after one Feynman parameter is integrated out:

\begin{equation}
\label{paramWMDM}
I_V^{A_1A_2A_3}=\int_0^1 F^{A_1A_2A_3}_V(x)dx .
\end{equation}
We  will first present the  $F^{A_1A_2A_3}_Z(x)$   functions necessary to calculate the AWMDM of a lepton.

\subsubsection{Anomalous weak magnetic dipole moment}
We  start by introducing the functions
\begin{equation}
f^{A_1A_2A_3}_1(x)=\frac{1}{\sqrt{Z^{A_1A_2A_3}(x)}}\arctan\left[\frac{X^{A_1A_2A_3}(x)}{\sqrt{Z^{A_1A_2A_3}(x)}}\right],
\end{equation}
and
\begin{equation}
f^{A_1A_2A_3}_2(x)=\log\left[Y^{A_1A_2A_3}(x)\right].
\end{equation}
For   diagram  (1)  we obtain
\begin{eqnarray}
F^{nVV}_Z(x)&=&x_\ell\Bigg[x \left(\left(3x(2-x)-2\right) x_Z+6 x^2-8 x+2\right)+
\frac{1}{2} x \left(\left(4 x^2-9 x+4\right) x_Z-8 x^2+12 x-4\right)f_2^{nVV}(x)\nonumber\\&+&
\Big(2 \Big(x^3 \left(x_Z-2\right) \left(x_n+x_{\ell }-1\right)+x^2 \left(2x_n(3-2 x_Z)-2 \left(x_Z-1\right) x_{\ell }+3 x_Z-8\right)\nonumber\\
&+&x \left(x_n \left(5 x_Z-4\right)+x_Z x_{\ell }-2\right)-2 x_n x_Z\Big)+
Z^{nVV}(x) \left(4 x^2 \left(x_Z-2\right)+x \left(12-9 x_Z\right)+4 \left(x_Z-1\right)\right)\Big)\nonumber\\&\times&f^{nVV}_1(x)\Bigg],
\end{eqnarray}
with
\begin{eqnarray}
X^{nVV}(x)&=&x x_Z,\\
Y^{nVV}(x)&=&(1-x) \left(x_n-x x_{\ell }\right)+x,\\
Z^{nVV}(x)&=&x_Z\left(4 Y^{nVV}(x)-x^2 x_Z\right),
\end{eqnarray}
where we have introduced the dimensionless variable $x_a=(m_a/m_V)^2$.

As for the contribution of   diagram (2),  the $F_Z^{Z'\ell\ell}$ function can be written as
\begin{equation}
\label{F^{Z'll}_Z}
F^{Z'\ell\ell}_Z=\left(g_L^{Z'\ell\ell}\right)^2 F_{Z-L}^{Z'\ell\ell}+\left(g_R^{Z'\ell\ell}\right)^2 F_{Z-R}^{Z'\ell\ell}+g_L^{Z'\ell\ell} g_R^{Z'\ell\ell}  F_{Z-LR}^{Z'\ell\ell},
\end{equation}
with
 \begin{eqnarray}
F_{Z-L}^{Z'\ell\ell}(x) &=&y_\ell\Bigg[2 x (2 x-1) g_L^{Z\ell\ell}-x (3 x-2) g_L^{Z\ell\ell}f^{Z'\ell\ell}_2(x)\nonumber\\
&-&2\Big(\left(2 x^2 y_{\ell }+6 x^2+6 \left(2-3x \right)+Z^{Z'\ell\ell}(x)\left(3x-2\right)\right)g_L^{Z\ell\ell} +2 x^2 y_{\ell } g_R^{Z\ell\ell}\Big)f^{Z'\ell\ell}_1(x)\Bigg],
\end{eqnarray}
\begin{equation}
F_{Z-R}^{Z'\ell\ell}(x)=F_{Z-L}^{Z'\ell\ell}(x)\left(g_L^{Z\ell\ell}\leftrightarrow g_R^{Z\ell\ell}\right),
\end{equation}
and
\begin{eqnarray}
F_{Z-LR}^{Z'\ell\ell}(x)&=&y_\ell
\left(g_L^{Z\ell\ell}+g_R^{Z\ell\ell}\right)\Bigg[2 (1-2 x) x+x (3 x-2)f^{Z'\ell\ell}_2(x)\nonumber\\&+&
2\Big(2x^2(2y_{\ell }+1)-14 x+Z^{Z'\ell\ell}(x)(3x-2)+12\Big)f^{Z'\ell\ell}_1(x)\Bigg],
\end{eqnarray}
with  $y_a=(m_a/m_{Z'})^2$ and
\begin{eqnarray}
X^{Z'\ell\ell}(x)&=& xy_Z,\\
Y^{Z'\ell\ell}(x)&=&x^2 y_{\ell }-x+1,\\
Z^{Z'\ell\ell}(x)&=&y_Z \left(4 Y^{Z'\ell\ell}(x)-x^2 y_Z\right).
\end{eqnarray}
We now present the contribution of diagram (3):
\begin{eqnarray}
F^{Vn n}_Z(x)&=&x_\ell\Bigg[2 x (2 x-1)+
(2-3 x) xf^{Vn n}_2(x)\nonumber\\&+&\Big(4 x^2
\left(x_n-2 x_{\ell }-3\right)+4x \left(2 (x_{\ell }- x_n)+9\right)+
2Z^{Vn n}(x) \left(2-3x\right)-24\Big)f^{Vn n}_1(x)\Bigg],
\end{eqnarray}
with
\begin{eqnarray}
X^{Vn n}(x)&=& xx_Z,\\
Y^{Vn n}(x)&=&x \left(x_n-x_{\ell }-1\right)+x^2 x_{\ell }+1,\\
Z^{Vn n}(x)&=&x_Z\left(Y^{Vnn}(x)-x^2 x_Z\right).
\end{eqnarray}
Finally, the loop function arising from diagram (4)  is given by

\begin{eqnarray}
F^{V\nu N}_Z(x)&=&\frac{x_\ell}{x_Z}\Bigg[4 x (2 x - 1) x_Z + 2 x  x_N \left(f^{V\nu N}_{21}(x)-f^{V\nu N}_{22}(x)\right)+ x (2 - 3 x)  x_Z \left(f^{V\nu N}_{21}(x)+f^{V\nu N}_{22}(x)\right) \nonumber\\&+&
   2 \sqrt{x_Z} \Big((x - 2) x x_N x_Z - (x - 2) x_N^2 - (3 x - 2) Z^{Vn n}(x) -
      2 (x - 1) x_Z (2 x x_\ell + 3 x - 6)\Big)\nonumber\\&\times&\left(f^{V\nu N}_{11}(x)+f^{V\nu N}_{12}(x)\right)\Bigg],
\end{eqnarray}
where the functions $f^{A_1A_2A_3}_{1a}(x)$ and $f^{A_1A_2A_3}_{2a}(x)$ are obtained from $f^{A_1A_2A_3}_{1}(x)$ and $f^{A_1A_2A_3}_{2}(x)$ after the replacement  $X^{A_1A_2A_3}(x)\to X_a^{A_1A_2A_3}(x)$ and $Y^{A_1A_2A_3}(x)\to Y_a^{A_1A_2A_3}(x)$, respectively.
In addition
\begin{eqnarray}
X^{V\nu N}_1(x)&=& x x_Z-x_n,\\
X^{V\nu N}_2(x)&=& x x_Z+x_n,\\
Y^{V\nu N}_1(x)&=&(x-1) \left(x x_{\ell }-1\right),\\
Y^{V\nu N}_2(x)&=&x \left(x_n-x_{\ell }-1\right)+x^2 x_{\ell }+1,\\
Z^{V\nu N}(x)&=&4Y^{V\nu N}_2(x)-x^2 x_Z.
\end{eqnarray}

We now turn to the parametric integrals for  the diagrams of Fig. \ref{WMDMSDiagrams}. For  diagram (1) we obtain
\begin{eqnarray}
F^{H\ell \ell}_Z(x)=-8x (2-x)w_{\ell}f_1^{H\ell \ell}(x),
\end{eqnarray}
with $w_a=(m_a/m_H)^2$ and
\begin{eqnarray}
X^{H\ell\ell}(x)&=&x w_Z,\\
Z^{H\ell \ell}(x)&=&w_Z\left((4(1-x(1-x w_\ell^2))-w_Z x^2\right).
\end{eqnarray}
Also, for the contribution of the pseudoscalar $\eta$ we obtain
\begin{eqnarray}
F^{\eta\ell \ell}_Z(x)=8 x^2z_{\ell}f_1^{\eta\ell \ell}(x),
\end{eqnarray}
with $z_a=(m_a/m_\eta)^2$, $X^{\eta\ell\ell}(x)=X^{H\ell\ell}(x)\left[w_a\to z_a\right]$, and $Z^{\eta\ell\ell}(x)=Z^{H\ell\ell}(x)\left[w_a\to z_a\right]$.

As for the contribution of  diagram (2) and the one obtained by exchanging the internal gauge boson with the scalar boson,  it is  as follows

\begin{eqnarray}
F^{\ell HV}_Z(x)&=&\frac{\sqrt{x_\ell}}{\sqrt{x_Z}}\Bigg[4 x (1-2 x) x_Z+\left((1-2 x) x_H+2 x-3\right)\left(f^{\ell HV}_{21}(x)-f^{\ell HV}_{22}(x)\right)\nonumber\\&+&(3 x-2) x x_Z\left(f^{\ell HV}_{21}(x)+f^{\ell HV}_{22}(x)\right)-2\sqrt{x_Z}\Big(x^2 x_Z \left(5 x_H+2 x_Z-28 x_{\ell }+5\right)
\nonumber\\&-&x \left(x_H \left(3 x_Z-4\right)+2 x_H^2-20 x_Z x_{\ell }+5 x_Z+2\right)+x_H(x_H-4)+3 x^3 x_Z \left(4 x_{\ell }-x_Z\right)-4 x_Z x_{\ell }+3\Big)\nonumber\\&\times&\left(f^{\ell HV}_{11}(x)-f^{\ell HV}_{12}(x)\right)
\Bigg],
\end{eqnarray}
where
\begin{eqnarray}
X^{\ell HV}_1(x)&=&x_H-x x_Z-1,\\
X^{\ell HV}_2(x)&=&x_H+x x_Z-1,\\
Y^{\ell HV}_1(x)&=& x^2 x_{\ell }-2 x x_{\ell }+x_{\ell }+x,\\
Y^{\ell HV}_2(x)&=& x^2 x_{\ell }-2 x x_{\ell }+x_{\ell }+x x_H,\\
Z^{\ell HV}(x)&=&2 x_H \left(x x_Z+1\right)-x_H^2+x^2 x_Z \left(4 x_{\ell
   }-x_Z\right)+x x_Z \left(2-8 x_{\ell }\right)+4 x_Z
   x_{\ell }-1.
\end{eqnarray}

\subsubsection{Anomalous magnetic dipole moment}

For completeness we present the contributions to the AMDM, which can be obtained from  the AWMDM  results after  taking the limit $m_Z\to 0$  and substituting the $Z$ coupling constants by the photon ones.  The $f^{A_1A_2A_3}_\gamma$ coefficients of Eq. (\ref{a_ell}) are presented in Table \ref{f^{A_1A_2A_3}}, whereas the respective loop integrals are of the form of (\ref{paramWMDM}).

As far as   Fig. \ref{WMDMGDiagrams} is concerned,  there are only contributions from diagrams (1) and (2), but with the external $Z$ gauge boson replaced by the photon. The contribution of diagram (1) can be written as
\begin{eqnarray}
\label{GVtauN}
F^{nVV}_\gamma(x)&=&-\frac{x x_{\ell }}{Y^{nnV}(x)} \left(x^2 \left(x_n+x_{\ell }+2\right)-x \left(3
   x_n+x_{\ell }-2\right)+2 x_n\right).
\end{eqnarray}
As for  the contribution of diagram (2),  the $F_\gamma^{Z'\ell\ell}$ function is given by an analogous
 expression to Eq. (\ref{F^{Z'll}_Z}) but now $F_{\gamma-R}^{Z'\ell\ell}(x)=F_{\gamma-L}^{Z'\ell\ell}(x)$, with
\begin{eqnarray}
\label{GZ'tautau}
F_{\gamma-L}^{Z'\ell\ell}(x)=- \frac{2 x y_{\ell }}{Y^{Z'\ell\ell}(x)}\left(x^2 \left(y_{\ell }+1\right)-3
   x+2\right),
\end{eqnarray}
and
\begin{eqnarray}
F^{Z'\ell\ell}_{\gamma-LR}(x)&=&\frac{ 4 x y_{\ell }}{Y^{Z'\ell\ell}(x)} \left(x^2 y_{\ell }-2 (x-1)\right).
\end{eqnarray}

In the limit of small $x_\ell$ the integration of the above functions  is straightforward and one obtains
\begin{eqnarray}
\label{GVtauN-1}
I^{nVV}_\gamma&=&-\frac{x_\ell}{6 (1 - x_n)^4} \left((1 - x_n) (10 - x_n (33 + x_n ( 4 x_n-45))) +
     18 x_n^3 \log(x_n)\right),
\end{eqnarray}
\begin{eqnarray}
\label{GZ'tautau-1}
I_{\gamma-LR}^{Z'\ell\ell}=-3I^{Z'\ell\ell}_{\gamma-L}= 4y_\ell.
\end{eqnarray}

Finally, there are also contribution of the scalar bosons $H$ and $\eta$ arising from the diagram (1) of Fig. \ref{WMDMSDiagrams}, with the $Z$ gauge boson replaced by the photon. The respective $F^{A_1A_2A_3}_\gamma (x)$ functions can be written as
\begin{equation}
F^{H\ell\ell}_\gamma(x)=\frac{2x^2(2-x)w_{\ell}}{w_{\ell} x^2+1-x},
\end{equation}
and
\begin{equation}
F^{\eta\ell\ell}_\gamma(x)=\frac{2x^3z_{\ell}}{z_{\ell}x^2+1-x}.
\end{equation}

All of the above results agree with previous calculations of the AMDM of a lepton (see for instance \cite{Leveille:1977rc,Jegerlehner:2009ry}).

\subsection{Passarino-Veltman scalar functions}
We now present the results for the AWMDM and AMDM of a lepton in terms of   Passarino-Veltman scalar functions.

\subsubsection{Anomalous weak magnetic dipole moment}
We first introduce the following ultraviolet finite functions given  in terms of two-point Passarino-Veltman scalar integrals:
\begin{eqnarray}
\Delta_1 &=& B_0(0, m_N^2, m_V^2)-B_0(m_\ell^2, m_N^2, m_V^2),\\
\Delta_2 &=& B_0(m_Z^2, m_V^2, m_V^2) -B_0(m_\ell^2, m_N^2, m_V^2),\\
\Delta_3 &=& B_0(0, m_\ell^2, m_{Z'}^2)-B_0(m_\ell^2, m_\ell^2, m_{Z'}^2) ,\\
\Delta_4&=&B_0(m_Z^2, m_\ell^2, m_\ell^2)-B_0(m_\ell^2, m_\ell^2, m_{Z'}^2),\\
\Delta_5 &=&B_0(m_Z^2, m_N^2, m_N^2) -  B_0(m_\ell^2, m_N^2, m_V^2),\\
\Delta_7&=&B_0(m_\ell^2, 0, m_V^2) -B_0(m_Z^2, 0, m_N^2),\\
\Delta_8&=&B_0(m_Z^2, 0, m_N^2) - B_0(0, 0, m_V^2),
\end{eqnarray}
and use a shorthand notation for the following three-point scalar functions
\begin{eqnarray}
 C_1&=&m_V^2 C_0(m_\ell^2, m_\ell^2, m_Z^2, m_V^2, m_N^2, m_V^2) ,\\
C_2&=&m_{Z'}^2C_0(m_\ell^2, m_\ell^2, m_Z^2, m_\ell^2, m_{Z'}^2, m_\ell^2),\\
C_3&=&m_V^2C_0(m_\ell^2, m_\ell^2, m_Z^2, m_N^2, m_V^2, m_N^2),\\
C_4&=&m_V^2 C0(m_\ell^2, m_\ell^2, m_Z^2, m_N^2, m_V^2, 0) .
\end{eqnarray}

The $I_Z^{A_1A_2A_3}$ loop functions arising from the diagrams of Fig. \ref{WMDMGDiagrams} are given as follows. For diagram (1) we obtain

\begin{eqnarray}
I_{Z}^{nVV}&=&\frac{x_\ell}{2\left(x_Z-4 x_{\ell}\right)^2}\Bigg[ \left(4 x_{\ell}-x_Z\right) \left(\left(x_n+x_{\ell}\right) \left(x_Z-2\right)-4\right)-\frac{1}{x_\ell} \left(x_n-1\right) \left(4 x_{\ell}-x_Z\right)  \left(\left(x_n+x_{\ell}\right) \left(x_Z-2\right)-4\right)\Delta_1\nonumber\\ &-& \Big(x_{\ell} \left(x_Z\left(x_Z+8\right)-60\right)+2\left(x_n\left( 3x_n-4 x_{\ell}\right)+ x_{\ell}^2\right) \left(x_Z-2\right)+x_n \left(x_Z\left(5 x_Z-16 \right)-12\right)-4 \left(x_Z\left(x_Z-3\right)-6\right)\Big)\Delta_2\nonumber\\ &+&
2  \Big(x_{\ell}^2 \left(24-5 x_n \left(x_Z-2\right)-x_Z\left( x_Z+4\right)\right)x_{\ell} \left(7 x_n^2 \left(x_Z-2\right)+x_n \left(3 x_Z^2-14 x_Z+24\right)+3 x_Z^2-x_Z-34\right)\nonumber\\&-&3 x_n^3 \left(x_Z-2\right)+2 x_n^2 \left(7-2 x_Z\right) x_Z-x_n \left(x_Z^3-7 x_Z^2+9 x_Z+18\right)+x_{\ell}^3 \left(x_Z-2\right)-2 x_Z^2+4 x_Z+12\Big)C_1\Bigg],
\end{eqnarray}
whereas the loop function arising from diagram (2) is given by a similar expression to Eq. (\ref{F^{Z'll}_Z}):

\begin{equation}
\label{I^{Z'll}_Z}
I^{Z'\ell\ell}_Z=\left(g_L^{Z'\ell\ell}\right)^2 I_{Z-L}^{Z'\ell\ell}+\left(g_R^{Z'\ell\ell}\right)^2 I_{Z-R}^{Z'\ell\ell}+g_L^{Z'\ell\ell} g_R^{Z'\ell\ell}  I_{Z-LR}^{Z'\ell\ell},
\end{equation}
where
\begin{eqnarray}
I_{Z-L}^{Z'\ell\ell}&=&-\frac{2 y_{\ell} }{\left(y_Z-4 y_{\ell}\right)^2}\Bigg[ \left(4 y_{\ell}-y_Z\right) \left(g_L^{Z\ell\ell} \left(y_{\ell}+2\right)+g_R^{Z\ell\ell} y_{\ell}\right)+\frac{1}{y_{\ell}}
 \left(y_{\ell}-1\right) \left(4 y_{\ell}-y_Z\right) \left(g_L^{Z\ell\ell} \left(y_{\ell}+2\right)+g_R^{Z\ell\ell} y_{\ell}\right)\Delta_3\nonumber\\ &-&
\left(g_L^{Z\ell\ell} \left(y_{\ell}\left(4y_{\ell}-\left(y_Z+34\right) \right)+2 \left(5 y_Z+6\right)\right)+g_R^{Z\ell\ell} y_{\ell} \left(4 y_{\ell}-y_Z+6\right)\right)\Delta_4\nonumber\\ &+&
2 \left(4 y_{\ell}-y_Z-3\right) \left(g_L^{Z\ell\ell} \left(2-7 y_{\ell}+2 y_Z\right)+g_R^{Z\ell\ell} y_{\ell}\right)C_2\Bigg],
\end{eqnarray}

\begin{eqnarray}
I_{Z-R}^{Z'\ell\ell}=I_{Z-L}^{Z'\ell\ell}\left(g_L^{Z\ell\ell}\leftrightarrow g_R^{Z\ell\ell}\right),
\end{eqnarray}
and
\begin{eqnarray}
I_{Z-LR}^{Z'\ell\ell}&=&\frac{2 y_{\ell}\left(g_L^{Z\ell\ell}+g_R^{Z\ell\ell}\right) }{\left(y_Z-4 y_{\ell}\right)^2}\Bigg[y_{\ell} \left(4 y_{\ell}-y_Z\right)+
\left(y_{\ell}-1\right)  \left(4 y_{\ell}-y_Z\right)\Delta_3-
\Big( y_{\ell}\left(4y_{\ell}-\left(y_Z+10\right) \right)+4 y_Z\Big)\Delta_4\nonumber\\&-&2
\Big(y_{\ell}\left(12 y_{\ell}-\left(7 y_Z+5\right) \right)+y_Z \left(y_Z+2\right)\Big)C_2\Bigg].
\end{eqnarray}

As for diagram (3), the respective loop integral is
\begin{eqnarray}
I_{Z}^{Vn n}&=&\frac{ x_{\ell}}{\left(x_Z-4 x_{\ell}\right)^2}\Bigg[\left(x_{\ell}+2\right) \left(x_Z-4 x_{\ell}\right)+
\frac{1}{x_{\ell}} \left(x_n-1\right) \left(x_{\ell}+2\right) \left(x_Z-4 x_{\ell}\right)\Delta_1\nonumber\\ &+&
\left(x_{\ell} \left(2 x_n+x_Z-22\right)-2 \left(x_n \left(x_Z+6\right)-5 x_Z-6\right)+2 x_{\ell}^2\right)\Delta_5+
2  \Big(x_{\ell}^2 \left(2 x_n+x_Z+12\right)-x_{\ell} \left(2 x_n \left(x_Z-5\right)+x_n^2+8 x_Z+17\right)\nonumber\\&+&x_n^2 \left(x_Z+6\right)-x_n \left(7 x_Z+12\right)-x_{\ell}^3+2 \left(x_Z^2+4 x_Z+3\right)\Big)C_3\Bigg].
\end{eqnarray}

Finally, the loop function arising from diagram (4)  obeys

\begin{eqnarray}
I_{Z}^{V\nu n}&=&\frac{1}{\left(x_Z-4 x_{\ell}\right)^2 x_Z}\Bigg[2 x_{\ell} x_Z \left(x_{\ell}+2\right) \left(x_Z-4 x_{\ell}\right)+ x_Z\left(x_n-1\right) \left(x_{\ell}+2\right) \left(x_Z-4 x_{\ell}\right)\Delta_1\nonumber\\ &-&
\Big(2 x_{\ell}^3 \left(6 x_n+x_Z\right)+x_{\ell}^2 \left(\left(x_Z-18\right) x_Z-2 x_n \left(3 x_Z+4\right)\right)+x_{\ell} x_Z \left(20-4 x_n+9 x_Z\right)-2 x_Z^2\Big)\Delta_6\nonumber\\ &+&
\Big( 2 x_{\ell}^3 \left(x_Z-6 x_n\right)+x_{\ell}^2 \left(8 x_n \left(x_Z+1\right)+\left(x_Z-26\right) x_Z\right)+ x_Z x_{\ell}\left(4-2 x_n \left(x_Z+4\right)+11 x_Z\right)+2 x_Z^2\Big)\Delta_7\nonumber\\&-&
x_{\ell} \Big(2 x_{\ell}^2 \left(x_Z-6 x_n\right)+x_{\ell} \left(8 x_n \left(x_Z+1\right)+\left(x_Z-22\right) x_Z\right)-2 x_Z \left(x_n \left(x_Z+4\right)-5 x_Z-6\right)\Big)\Delta_8\nonumber\\ &+&
2 x_{\ell} \Big(x_{\ell}^2 \left(2 x_n \left(x_Z-3 x_n\right)+2 x_Z \left(x_Z+12\right)\right)+x_{\ell} \left(x_n^2 \left(3 x_Z+4\right)-2 x_Z\left( x_n \left(x_Z-5\right)+ \left(8 x_Z+17\right)\right)\right)\nonumber\\&-&x_Z \left(x_n \left(7 x_Z+12\right)-2 x_n^2-4 \left(x_Z^2+4 x_Z+3\right)\right)-2 x_{\ell}^3 x_Z\Big)C_4\Bigg].
\end{eqnarray}
We now present the loop functions for the diagrams of Fig. \ref{WMDMSDiagrams}. We will use the following additional Passarino-Veltman scalar functions

\begin{eqnarray}
\Delta_9&=&B_0(0,m_H^2,  m_\ell^2) -B_0(m_Z^2, m_\ell^2, m_\ell^2),\\
\Delta_{10}&=&B_0(m_\ell^2,m_H^2,  m_\ell^2) -B_0(m_Z^2, m_\ell^2, m_\ell^2),\\
C_5&=&m_H^2C_0(m_\ell^2, m_\ell^2, m_Z^2, m_\ell^2, m_H^2, m_\ell^2).
\end{eqnarray}
For diagram (1) we  obtain for the contribution of the SM Higgs boson $H$
\begin{eqnarray}
I_{Z}^{H\ell\ell}&=&\frac{2}{\left(w_Z-4 w_{\ell}\right)^2}\Bigg[w_{\ell} \left(w_Z-4 w_{\ell}\right)-
 \left(w_{\ell}-1\right) \left(4 w_{\ell}-w_Z\right)\Delta_9\nonumber\\ &+&
 \left(16 {w_{\ell}}^2-2 w_{\ell} \left(2 w_Z+5\right)+w_Z\right)\Delta_{10}-
6 w_{\ell} \left(4 w_{\ell}-w_Z-1\right)C_5\Bigg],
\end{eqnarray}
and for the contribution of the new pseudoscalar $\eta$
\begin{eqnarray}
I_{Z}^{\eta\ell\ell}&=&\frac{2}{\left(z_Z-4 z_{\ell}\right)^2}\Bigg[z_{\ell} \left(z_Z-4 z_{\ell}\right)-
 \left(z_{\ell}-1\right) \left(4 z_{\ell}-z_Z\right)\Delta_{11}+
\left(z_Z-10 z_{\ell}\right)\Delta_{12}-2 z_{\ell} \left(4 z_{\ell}-z_Z-3\right)\Big)C_6\Bigg],
\end{eqnarray}
where $\Delta_{11}$, $\Delta_{12}$, and $C_6$ are obtained from  $\Delta_{9}$, $\Delta_{10}$ and $C_5$, respectively, after the replacement $m_H\to m_\eta$.

As for diagram (2) of Fig. \ref{WMDMSDiagrams}, it yields

\begin{eqnarray}
I_{Z}^{\ell HV}&=&\frac{2\sqrt{x_{\ell }}}{ \sqrt{x_Z} \left(4 x_{\ell }-x_Z\right)}\Bigg[  x_{\ell } \left(x_{\ell}-x_H\right)\Delta_{13}+
 \left(x_{\ell }-1\right) x_{\ell }\Delta_{14}
-4  x_{\ell }\Delta_{15}-2 x_Z\Delta_{16}+
2\left(2 x_ {\ell } \left(x_H+x_Z-1\right)-x_H x_Z\right)C_ 7\Bigg],
\end{eqnarray}
with
\begin{eqnarray}
\Delta_{13}&=&B_0(0,m_H^2,  m_\ell^2) -B_0(0, m_H^2, m_V^2),\\
\Delta_{14}&=&B_0(0, m_H^2, m_V^2) -B_0(0, m_\ell^2, m_V^2),\\
\Delta_{15}&=&B_0(m_\ell^2,m_H^2,  m_\ell^2) -B_0(m_\ell^2, m_\ell^2, m_V^2),\\
\Delta_{16}&=&B_0(m_\ell^2, m_\ell^2, m_V^2) -B_0(m_Z^2, m_H^2, m_V^2),\\
C_7&=&m_V^2C_0(m_\ell^2, m_\ell^2, m_Z^2, m_H^2, m_\ell^2, m_V^2).
\end{eqnarray}

\subsection{Anomalous magnetic dipole moment}

For the loop integrals of the contributions to the AMDM of the diagrams analogue to those of Fig. \ref{WMDMGDiagrams}, but with the external $Z$ boson replaced by the photon, we have for diagram (1)

\begin{eqnarray}
I_\gamma^{nVV}&=&-\frac{1}{8x_\ell}\Bigg[\left(x_n x_{\ell} \left(5 x_{\ell}+12\right)-7
   x_n^2 x_{\ell}-9 x_n+3 x_n^3-x_{\ell}
   \left(\left(x_{\ell}-12\right) x_{\ell}+17\right)+6\right)C'_1 \nonumber\\&+&2 \left(x_n-1\right)
   \left(x_n+x_{\ell}+2\right) \Delta _1+ \left(\left(15-4
   x_n\right) x_{\ell}+3 \left(x_n+x_n^2-2\right)+x_{\ell}^2\right)\Delta' _2-2 x_{\ell}
   \left(x_n+x_{\ell}+2\right)\Bigg],
\end{eqnarray}
where the primed scalar functions $\Delta'_i$ and $C'_i$ are obtained from the unprimed ones by setting $m_Z=0$. We note that all the  three-point functions  $C'_i$ appearing in the AMDM are of the generic type
$C_0(m_A^2,m_A^2,0,m_B^2,m_C^2,m_B^2)$, which can be written in terms of two-point scalar functions as follows \cite{Devaraj:1997es}

\begin{eqnarray}
C_0(m_A^2,m_A^2,0,m_B^2,m_C^2,m_B^2)=\frac{1}{\lambda(m_A^2,m_B^2,m_C^2)}
\Big( \left(m_A^2-m_B^2-m_C^2\right)\Delta _B-2 m_C^2\Delta_C+2\left(m_A^2-m_B^2+m_C^2\right)\Big),
\end{eqnarray}
with $\lambda(x,y,z)=(x-y-z)^2-4yz$ and $\Delta _X=B_0(0,m_X^2,m_X^2)-B_0(m_A^2,m_B^2,m_C^2)$.

As for  diagram (2) we obtain a similar expression to Eq. (\ref{I^{Z'll}_Z}),

where
\begin{eqnarray}
I_{\gamma-R}^{Z'\ell\ell}&=&I_{\gamma-L}^{Z'\ell\ell}=-\frac{1}{4y_\ell}\Bigg[ \left(3 y_{\ell}-1\right) \left(4 y_{\ell}-3\right)C'_2+ 2\left(1- {y_{\ell}}^2\right)\Delta_3+  \left(y_{\ell}-3\right) \left(2 y_{\ell}-1\right)\Delta' _4-2
   y_{\ell} \left(y_{\ell}+1\right)\Bigg],
\end{eqnarray}
and
\begin{eqnarray}
I^{Z'\ell\ell}_{\gamma-LR}&=&-\frac{1}{2}\Bigg[\left(5-12 y_{\ell}\right)C'_2+2 \left(y_{\ell}-1\right)\Delta _3 + \left(5-2y_{\ell}\right)\Delta' _4+2 y_{\ell}\Bigg].
\end{eqnarray}
Finally, the diagram (1) of Fig \ref{WMDMSDiagrams} with the $Z$ replaced by the photon yields the following loop functions
\begin{eqnarray}
I^{H\ell\ell}_{\gamma}&=&-\frac{1}{4w_{\ell}}\Bigg[2w_{\ell}+
   2\left(w_{\ell}-1\right)\Delta' _9+\left(5-8 w_{\ell}\right) \Delta'_{10}+3\left(4 w_{\ell}-1\right)C'_5 \Bigg],
\end{eqnarray}
and
\begin{eqnarray}
I^{\eta\ell\ell}_{\gamma}&=&-\frac{1}{4z_{\ell}}\Bigg[2x_{\ell}+ 2\left(x_{\ell}-1\right)\Delta' _{11}+5 \Delta'_{12}+\left(4 x_{\ell}-3\right)C'_6 \Bigg].
\end{eqnarray}
Again the primed scalar functions $\Delta'_i$ and $C'_i$ are obtained from the unprimed ones by setting $m_Z=0$.

\section{Decay $\ell_i\to \ell_j\gamma$}
\label{litoljgammaappendix}
In this appendix we present the amplitude for the $\ell_i\to \ell_j\gamma$ decay both in terms of parametric integrals and Passarino-Veltman scalar functions. The  contributions arise from the Feynman diagrams of Fig. \ref{litoljgammadiagram}. The decay width is given in Eq. (\ref{litoljgammadecaywidth}). We have obtained the  $f_L(x_{k} )$  function [$x_k=(m_{N_k}/m_V)^2$] appearing in Eq. (\ref{FL}) via the Feynman parameters technique using the approximation of massless final lepton $m_{\ell_j}=0$. We first define the function

\begin{eqnarray}
f(x;z)&=&\frac{2}{x x_{\ell_i}} \left(2(1-
   x_{\ell_i})+z+x \left(\left(x
   \left(z-1\right)+z+3\right)
   \left(z-1\right)-x_{\ell_i}
   \left(z-2\right)\right)\right)\log \left(\frac{1+x \left((x-1)
   x_{\ell_i}+z -1\right)}{1+x(z-1)}\right)\nonumber\\&+&(x-1) \left(\frac{x_{\ell_i} (x-1) x^2 \left((2 x-1)
   x_{\ell_i}+z-1\right)}{x \left((x-1)
   x_{\ell_i}+z-1\right)+1}-2
   \left(x+1\right)z\right)\simeq
   \frac{(1-x) \left((z+2) x^2+(z-6)
   x+4\right)}{x
   (z-1)+1}x_{\ell_i}+O\left(x_{\ell_i}^2\right),
\end{eqnarray}
where $x_{\ell_i}=(m_{\ell_i}/m_V)^2$ ($V=W,X$), whereas  the $f_L$ function is given as

\begin{equation}
\label{fLPF}
f_L(x_k)=\int_0^1 dx f(x;x_k)\simeq \frac{1}{6 (x_k-1)^4} \left(18 \log (x_k) x_k^3+(x_k-1)
   (x_k (x_k (4 x_k-45)+33)-10)\right)x_{\ell_i}+O\left(x_{\ell_i}^2\right).
\end{equation}
For the sake of completeness we also include the amplitude in terms of Passarino-Veltman scalar functions. The result  reads as

\begin{eqnarray}
\label{fLPV}
f_L(x_k)&=&\frac{
   \left(2(1+x_k)-x_{\ell_i}
   \right)x_k}{x_{\ell_i}}+\frac{2 x_k
   \left((1+x_k)x_k-2-x_{\ell_i}
   \left(x_k-2\right)\right)}{x_{\ell_i}
   \left(x_k-1\right)}B_{\ell_i {N_k}V}+2 \left(2(1- x_{\ell_i})+x_k\right) C_{\ell_i {N_k}V},
\end{eqnarray}
where we have defined
\begin{eqnarray}
B_{\ell_i{N_k}V}&=&\left(1-x_k\right)B_0(m_{\ell_i}^2,m_{N_k}^2,m_{N_k}^2)+x_k B_0(0,m_{N_k}^2,m_{N_k}^2)-B_0(0,m_V^2,m_V^2),\nonumber\\
C_{\ell_i {N_k}V}&=&m_V^2C_0(m_{\ell_i}^2, 0, 0, m_{N_k}^2, m_V^2, m_V^2).
\end{eqnarray}

\twocolumn

\end{document}